\documentclass[sts]{imsart}
\RequirePackage{amsthm,amsmath,amsfonts,amssymb}
\RequirePackage[authoryear]{natbib}
\RequirePackage[colorlinks,citecolor=blue,urlcolor=blue]{hyperref}
\RequirePackage{graphicx}

\startlocaldefs
\theoremstyle{plain}
\newtheorem{axiom}{Axiom}

\newtheorem{theorem}{Theorem}[section]

\newtheorem{assumption}{Assumption}
\newtheorem{propo}[theorem]{Proposition}

\theoremstyle{definition}
\newtheorem{definition}[theorem]{Definition}

\usepackage{thmtools, thm-restate}

\newcommand{\e}[1]{\mathbb{E}\left[#1\right]  } 
\newcommand{\comment}[1]{%
   \tag{#1}
}
\usepackage{centernot}

\newcommand{\indep}{\mathrel{\mkern1mu\perp\mkern-9mu\perp\mkern1mu}}
\newcommand{\notindep}{\centernot\indep}

\endlocaldefs

\usepackage{xcolor}         
\usepackage[textwidth=3cm]{todonotes}

\usepackage{cleveref}
\usepackage{booktabs} 
\usepackage{longtable}
\usepackage{cancel}

\usetikzlibrary {arrows.meta} 
\tikzset{>={Latex[length=2mm]}}

\begin{document}

\begin{frontmatter}
\title{A Principled Approach for Defining and Comparing Variable Importance Measures}
\runtitle{A Principled Approach for Defining and Comparing Variable Importance Measures}


\begin{aug}
\author[A]{\fnms{Angel}~\snm{Reyero Lobo}
\ead[label=e1]{angel.reyero-lobo@inria.fr}},
\author[B]{\fnms{Pierre}~\snm{Neuvial}\ead[label=e2]{pierre.neuvial@math.univ-toulouse.fr}}
\and
\author[C]{\fnms{Bertrand}~\snm{Thirion}\ead[label=e3]{bertrand.thirion@inria.fr}
}


\address[A]{Angel Reyero Lobo is PhD candidate, Institut de Mathémathiques de Toulouse and Inria Saclay, France\printead[presep={\ }]{e1}.}

\address[B]{Pierre Neuvial is Research Director, Institut de Mathémathiques de Toulouse, France\printead[presep={\ }]{e2}.}

\address[C]{Bertrand Thirion is Research Director, Inria Saclay, CEA, Université Paris Saclay, France\printead[presep={\ }]{e3}.}

\end{aug}

\begin{abstract}
Variable importance measures (VIMs) aim to quantify the contribution of each input covariate to the predictability of a given output. With the growing interest in explainable AI, numerous VIMs have been proposed, many of which are heuristic in nature. This is often justified by the inherent subjectivity of the notion of importance. This raises important questions regarding usage: What makes a good VIM? How can we compare different VIMs?

In this paper, we address these questions by: (1) proposing an axiomatic framework that bridges the gap between variable importance and variable selection. This allows VIM methods to leverage the extensive literature on statistical guarantees developed for conditional independence testing. Our framework formalizes the intuitive principle that features providing no additional information should not be assigned importance, helping avoid false positives due to spurious correlations, which can arise with popular methods such as Shapley values; and (2) advocating for a general pipeline for constructing VIMs based on the definition of a theoretical index, its estimation, and the derivation of associated statistical guarantees. This pipeline clarifies the objective of various VIMs and thus facilitates meaningful comparisons. While this approach is natural from a statistical perspective, much of the literature has diverged from it, as we illustrate using the popular Total Sobol' Index. Finally, we apply this framework to an extensive set of VIMs, showing how practitioners can select and estimate indices aligned with their specific goals. Our results are supported by numerical experiments on simulated and real data.
\end{abstract}


\end{frontmatter}

\section{Introduction}

Global variable importance aims to assign a measure of relevance to each feature with respect to a target.
Since this relationship can be highly complex, machine learning (ML) models—proven effective at capturing such complexities in real-world scenarios—are often used as surrogates. 
%
This is at the core of scientific discovery in data-driven approaches that are currently pervasive. 

However, ML models are often opaque. While simple models like linear regression are fully interpretable through their coefficients, they rarely reflect the true data-generating process. Hence, there exists a trade-off between model complexity and interpretability \citep{Molnar2022}. To mitigate this, there has been growing interest in \textit{model-agnostic} variable importance measures (VIMs), see \cite{williamson2023general}.

Controlled variable selection has emerged as a method for filtering features with statistical guarantees \citep{candes2018}. 
Although variable selection and variable importance may appear to pursue the same goal, they have been treated separately in the literature. 
For instance, some variable importance measures aim to equitably distribute relevance across all features, while variable selection aims to identify a minimal subset of predictive variables \citep{scornet2022mda}. 
One of the reasons is that some popular variable importance measures are based on complex combinatorial computations and game-theoretic axioms that make them hard to interpret and not well suited for this goal \citep{LocoVSShapley}.

\begin{figure}[htbp]
    \centering
        \includegraphics[width=0.45\textwidth]{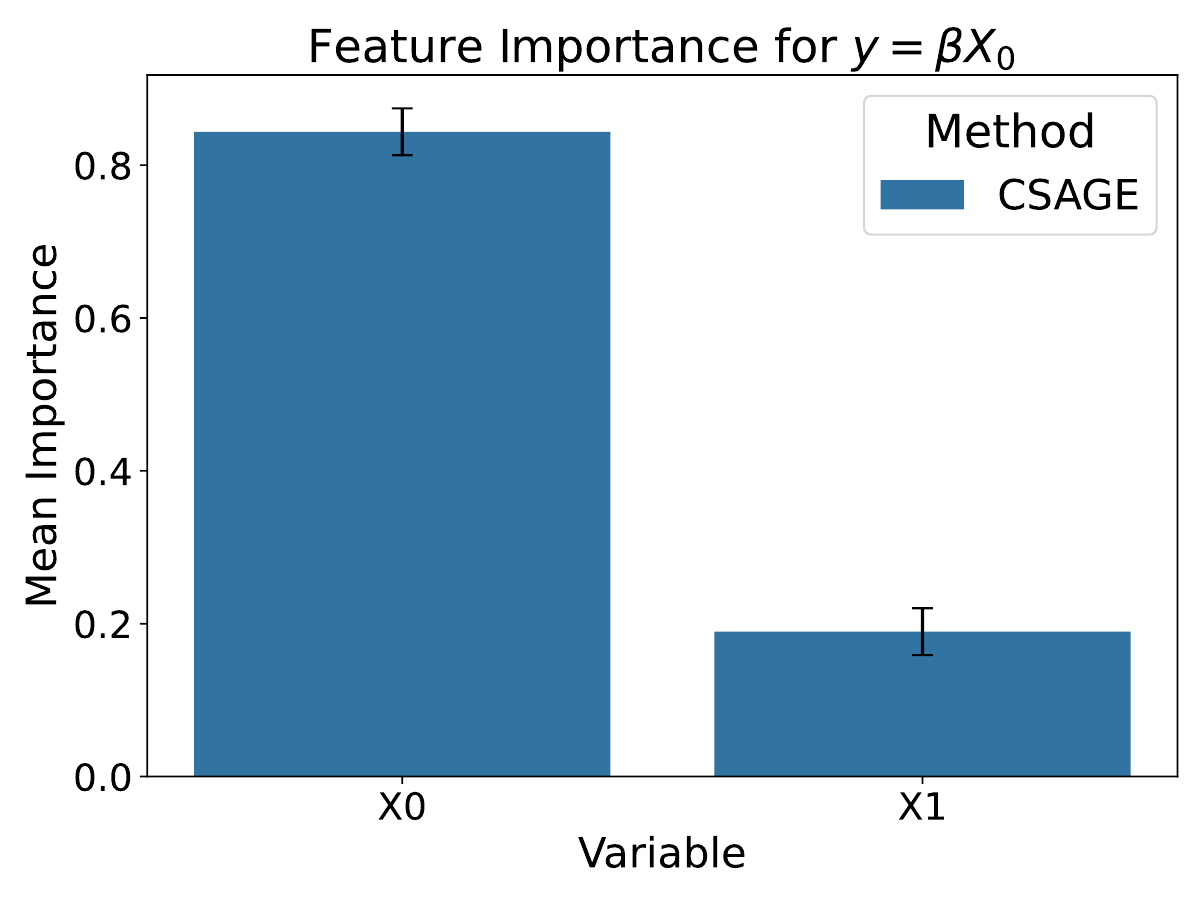}

    \caption{\textbf{Bar plot of the estimated importance using SAGE:} 
        The input consists of two Gaussian features correlated by $0.6$, with a linear output $y = \beta X_0$. 
        Left: importance of $X_0$; Right: importance of $X_1$. 
        Note that Shapley values assigns nonzero importance to $X_1$, even though it does not appear in the true model. 
        Importance was estimated using Gradient Boosting with $n = 10{,}000$ samples over 100 repetitions.
}
    \label{fig:motivating_ma}
\end{figure}

The most widely used importance measure is the Shapley value \citep{Shapley1953}, which, for global variable importance, is approximated by conditional SAGE \citep[cSAGE, ][]{covert2020understanding}, further discussed in Section \ref{subsec:Shap_not_minimal}. 
Figure~\ref{fig:motivating_ma} illustrates this approach in a simple linear setting with a single relevant covariate. Notably, Shapley values assign nonzero importance to irrelevant yet correlated features. This potentially leads to false discoveries and questions its applicability to scientific discovery.

Ideally, we want a definition of importance that considers the impact of missing a feature on predictive performance. This definition should align with the standard definitions of importance from \cite{covert2020understanding, molnar2025}. However, this view conflicts with current axiomatic frameworks \citep{Shapley1953, LocoVSShapley}. In response, we propose a \textbf{minimal axiom}: variables that do not provide any exclusive information should not be assigned importance. It bridges the gap between variable selection and importance by offering both statistical guarantees and rich information beyond binary decisions. In practice, variable importance measures are often used for feature selection by retaining the highest-ranked variables. However, this heuristic provides no control of selection errors and ignores the randomness arising from both the sampled data and the estimation procedure. Consequently, it is desirable to connect variable importance with principled variable selection methods that offer statistical guarantees by imposing a common null hypothesis. Our proposed axiom is \emph{minimal} in the sense that it corresponds to a necessary condition for null importance, allowing for flexibility to accommodate subjective definitions of importance.%

We conduct a comprehensive evaluation of state-of-the-art VIMs in the light of this axiom. 
Notably, Shapley values do not satisfy the proposed axiom, meaning that they may assign importance to unimportant features, as seen in Figure \ref{fig:motivating_ma}. 
%
%
%
%
%
%
In contrast, we show that many VIMs satisfy this property because the Bayes predictor does not depend on features that are conditionally independent of the response, as such features are not predictive. Moreover, at the empirical level, we expect that an \emph{implicit conditional filtering} mechanism emerges during model training, and thus the fitted model mainly relies on the truly informative features. This assumption is closely related to the intuition underlying the use of ML for scientific discovery: the trained model acts as a surrogate for the underlying data-generating distribution and can therefore provide insight into the importance of individual features in the Bayes predictor. We support this hypothesis through extensive experiments in both regression and classification settings using a wide range of models, including Gradient Boosting, Random Forests, Neural Networks, as well as more sophisticated ensemble methods such as Super Learner \citep{van2007super} and the recent TabICL \citep{qu2026tabiclv2}.

One substantial challenge in VIMs is the inconsistency across rankings produced by different methods, making it difficult to assess which features are truly important.
Therefore, it is necessary to classify them to enable insightful comparisons.
Existing categorizations mainly focus on the inference step and how the methods removes features \citep{covert_unifying}—grouping methods by whether they involve refitting, marginalizing, or perturbing features \citep{Ewald_2024}. However, this overlooks the fact that the model is just a surrogate for the data-generating function, which is unknown in practice and must be estimated. In practice, it is not any model, but the minimizer of an empirical loss. Moreover, variable importance estimation often involves quantities like conditional distributions or expectations, which are often ignored in the theoretical analysis \citep{covert2020understanding, Hooker2021, Ewald_2024}.
Furthermore, methods that use the same estimation procedure can yield different rankings, as demonstrated in our experiments comparing Permutation- and Conditional- feature importance —PFI and CFI, both perturbation-based— as well as mSAGE and cSAGE—both marginalization-based. 

To enable interpretable comparisons, we advocate a \textbf{principled approach for constructing VIMs}. This  approach begins by \textbf{(1) explicitly defining a theoretical importance index} (which should satisfy the minimal axiom) that aligns with user objectives. 
Then, it focuses on \textbf{(2) estimating} this index, considering desirable inference properties such as robustness to model misspecification \citep{lobo2025doublerobustnessconditionalfeature}. Finally, to enable reliable scientific discovery, we emphasize \textbf{(3) recovering statistical guarantees}, such as type-I error control. We argue that the uncertainty of the model should therefore be accounted for only at the last two stages, not in the definition of the theoretical index itself.

Next, we study existing importance indices in the light of this three-step approach: we explain their interpretations, compare different estimation strategies, and outline the corresponding statistical guarantees to guide practitioners in real-world applications.

The main contributions of this paper are:
\begin{itemize}
    \item to propose an interpretable, minimal axiom that unifies variable selection and importance.
    \item to introduce a principled approach for constructing VIMs and comparing them meaningfully.
    \item to apply the proposed methodology to an extensive set of VIMs, providing a new classification, distinct from recent presentation of the field.
    \item to illustrate using extensive numerical experiments with both real and synthetic datasets that the VIMs proven to satisfy the minimal axiom assign zero importance to the same null covariates, and the newly proposed classification is consistent with the obtained importance rankings.
\end{itemize}

\subsection*{Setting and Notation}
Let $(X, Y) \sim P_{(X, Y)}$, where $P_{(X, Y)} \in \mathcal{M}$, $X \in \mathcal{X} \subset \mathbb{R}^p$ denotes the input, and $Y \in \mathcal{Y} \subset \mathbb{R}$ denotes the output. The pair $(X, Y)$ is sampled from a distribution $P_{(X, Y)}$ belonging to the model class $\mathcal{M}$. When there is no ambiguity, we write $P$ instead of $P_{(X, Y)}$.
Given a set $S \subset [p]$, we denote by $-S=[p]\backslash S$ the subset $[p]$ restricted of $S$. If $S=\{j\}$, when it is clear from the context, $j$ (resp. $-j$) denotes the subset $\{j\}$ (resp. $-\{j\}$). 
We denote by $X^S$ the coordinates of $X$ corresponding to the subset $S$. Similarly, we denote by $X^j$ the $j$-th coordinate of $X$. 

We define $\psi(j, P_{(X, Y)})$ as the importance index of the $j$-th covariate, $X^j$, under the distribution $P_{(X, Y)} \in \mathcal{M}$, for $j \in \{1, \ldots, p\}$. When the underlying distribution is clear from the context, we simply write $\psi(j)$.
We denote by $\mathcal{F}$ an arbitrary space of functions from $\mathcal{X}$ to $\mathcal{Y}$. We denote by $m\in\mathcal{F}$ the conditional expectation $m(X)=\e{Y\mid X}$.

We emphasize that no prior assumption is made on the function space $\mathcal{F}$. Thus, there is complete freedom in choosing the function $m\in \mathcal{F}$. However, in practice, inference challenges may arise when approximating $m$ using a machine learning model $\widehat{m}$, motivating the use of a model-agnostic approach to accommodate complex data-generating processes. We denote by $\widehat{m}_n$ the estimator that makes explicit the dependence on the training sample size $n$. This estimation issue is further discussed in Section \ref{sect:pipeline}. It is obtained by minimizing a loss function $\ell$.

Some procedures require evaluating a function $f$ on samples where the feature $X^j$ is replaced by another random variable $\widetilde{X}^j$, while the remaining covariates $X^{-j}$ are kept unchanged. For readability, instead of introducing a dedicated operator to construct the perturbed vector, we simply write $f(\widetilde{X}^j, X^{-j})$, with the understanding that the arguments are arranged in the appropriate order.

We also make the identifiability assumption that no input covariate is an exact function of the others. This ensures that the problem of assigning importance is well-defined—a standard assumption in the variable importance literature \citep[see][]{candes2018, LocoVSShapley}. If necessary, this can be achieved by prior filtering of repeated features or a choice of a representative of the group \citep{MEINSHAUSEN20131439}. Under this assumption, the set of important covariates
\[
\mathcal{S} := \left\{ j \in \{1, \ldots, p\} \,\middle|\, X^j \notindep Y \mid X^{-j} \right\}
\]
is well defined \citep{candes2018}. This set is known as the \emph{Markov blanket} of $Y$ and is the target of variable selection \citep{Shah_2020}. It is the smallest set of features that renders the output conditionally independent of all remaining features. 

In Section~\ref{sect:related_work}, we revisit the axiomatic approaches proposed in the literature. In Section~\ref{sect:axiom}, we introduce our new interpretable minimal axiom, designed to close the gap between variable importance and variable selection. Section~\ref{sect:pipeline} develops our principled framework for constructing and comparing Variable Importance Measures (VIMs). In Section~\ref{sect:examples}, we apply this framework to state-of-the-art VIMs and discuss whether they satisfy the minimal axiom, with a summary provided in Section~\ref{sect:overview}. Finally, Section~\ref{sec:exp} presents simulations and real data analyses illustrating our findings.

\section{Related work}\label{sect:related_work}

The notion of importance is inherently subjective, and formalizing what is desirable from a mathematical perspective can be challenging.

In this section, we begin by discussing the main axiomatic foundations that have been proposed in the literature. In Section \ref{subsec:shap_axioms}, we review the Shapley value axioms, and in Section \ref{subsec:verd_axioms}, we present the critique put forward by \cite{LocoVSShapley} along with their newly proposed axioms.

\subsection{Shapley values axioms}\label{subsec:shap_axioms}

\cite{Shapley1953} introduced an axiomatic framework for the fair allocation of a total payout among players in a cooperative game, where each player's share is determined by their contributions across all possible subsets of players. Building on this idea, \cite{covert2020understanding} applied the same framework to variable importance, interpreting the features as ``players'' and the ``game'' as the prediction task, where the value corresponds to the model's predictive performance. Indeed, it has been often proposed in the literature as a solution for handling dependent covariates \citep{OwenPrieur2017ShapleyDependent}.

Let $V : \mathcal{P}([p]) \to \mathbb{R}$ be a value function, where $\mathcal{P}$ denotes the power set operator and $S \in \mathcal{P}([p])$ represents a subset of indices. This value function assigns a measure of predictive power to each feature set $S$. It consists of the building blocks used to define the importance measure $\psi$, which aggregates importance across all subsets. For instance, \cite{covert2020understanding} propose defining $V(S)$ as the difference in the model loss when using the covariates in $S$ versus using no covariates (i.e., predicting by the mean).

The classical Shapley axioms are as follows:

\begin{axiom}[Efficiency]\label{ax:effic}
The total importance is fully distributed among all features: $\sum_j \psi(j, P) = V([p])$.
\end{axiom}

\begin{axiom}[Symmetry]
If two features $j$ and $k$ contribute equally to every subset, i.e., $V(S \cup \{j\}) = V(S \cup \{k\})$ for all $S$, then $\psi(j, P) = \psi(k, P)$.
\end{axiom}

\begin{axiom}[Dummy]\label{ax:dummy}
If a feature $j$ does not contribute to any subset, i.e., $V(S \cup \{j\}) = V(S)$ for all $S$, then $\psi(j, P) = 0$.
\end{axiom}

\begin{axiom}[Linearity]
If two value functions $V$ and $V'$ yield values $\psi(j, P)$ and $\psi'(j, P)$ respectively, then the value of $V + V'$ is $\psi(j, P) + \psi'(j, P)$.
\end{axiom}

This framework led to the now well-known Shapley values, which are the unique fulfilling this axiomatic framework. A common criticism of Shapley values is their computational burden due to the need to evaluate all possible feature subsets. However, in this work, we focus on conceptual critiques rather than computational concerns.

\cite{LocoVSShapley} criticized Shapley values from both inferential and conceptual perspectives. Inferentially, many combinations of features may lead to poor estimation, making it questionable to rely on all possible subsets. 
Conceptually, due to the complex weighted combinations, these values are difficult to interpret, as their numerical magnitudes cannot be directly associated with a tangible property, such as the percentage of explained variance or the effect of a feature within the model.


Moreover, the efficiency axiom \eqref{ax:effic} imposes an additivity constraint, which may be undesirable when variables are correlated or interact non-linearly \citep{kumar}. Such assumptions may distort the attribution of importance. 

Due to the strict nature of the dummy axiom \eqref{ax:dummy}, when using the value function proposed by \cite{covert2020understanding}, 
a feature is assigned zero importance only in the very specific case where it is completely independent of both the target and all other features. Hence, Shapley values may assign nonzero importance to irrelevant covariates, as seen in Figure \ref{fig:motivating_ma}. This undermines their use in feature selection, and therefore scientific discovery. 

\subsection{Verdinelli \& Wasserman axioms}\label{subsec:verd_axioms}

\cite{LocoVSShapley} tackled the issue of \emph{correlation distortion}—the phenomenon where the importance of a feature is underestimated due to its correlation with other inputs. They proposed a new axiomatic foundation for regression settings, explicitly aligning with the interpretation of importance as predictive power and aiming to mitigate the effects of correlation:

\begin{axiom}[Functional Dependence]\label{ax:func_dep}
$\psi(j, P) = 0$ if and only if $m(X)$ is not a function of $X^j$.
\end{axiom}

\begin{axiom}[Correlation-Free]\label{ax:corr_free}
$\psi(j, P) = \psi(j, p(Y \mid X) \, p_j(X^j) \, p_{-j}(X^{-j}))$, where $p_j$ and $p_{-j}$ denote the marginals of $X^j$ and $X^{-j}$, respectively.
\end{axiom}

\begin{axiom}[Linear Agreement]\label{ax:lin_agree}
If $m(X) = \sum_j \beta_j X^j$, then $\psi(j, P) = \beta_j^2$.
\end{axiom}

The first axiom is intuitive: a feature is important if the prediction function $m(X)(:=\e{Y\mid X})$ depends on it. The second axiom aims to eliminate correlation distortion by requiring that importance depends only on the marginals of $X^j$ and $X^{-j}$, and not on their joint dependence structure. The third axiom ensures agreement with the coefficients in a linear model, grounding the measure in familiar settings.

In Section \ref{sect:axiom}, we propose a generalization of the functional dependence axiom and argue that the correlation-free axiom may not be universally appropriate, as its necessity depends on one’s subjective view on what variable importance should capture.

Additionally, \cite{LocoVSShapley} introduced several more loosely defined axioms motivated by computational and inferential considerations. However, these are not fully satisfied by their proposed VIM, a point we further discuss in Section \ref{sec:TSI}.

\section{Axiomatic Framework for Variable Importance Measures}\label{sect:axiom}

The notion of variable importance is inherently subjective. Consequently, an axiomatic framework—such as those previously proposed—may be overly restrictive or tailored to specific objectives, making it difficult to generalize. For instance, the correlation-free requirement (Axiom \ref{ax:corr_free}) may not be universally desirable. Depending on the objective, a practitioner may prefer to decrease the importance of a feature when its information is largely redundant with that of other features, while increasing the importance of features that provide genuinely new information.

For example, in house-price prediction, the number of rooms and surface area are typically correlated and may contain overlapping information. If the goal is to measure unique contributions, it may be desirable to discount their importance due to this redundancy, rather than maintaining or inflating their importance. In contrast, neighborhood may provide complementary information not captured by structural characteristics and could therefore deserve greater importance. Conversely, if the goal is purely predictive utility, such a correlation-based discount may be undesirable, since both correlated features can still be useful predictors. Thus, whether correlation-induced changes in importance constitute a distortion depends on the objective underlying the importance measure.

However, there is a consensus in the literature that feature importance should be at least related to a feature’s predictive power \citep{covert2020understanding, LocoVSShapley, Ewald_2024}. Notably, \cite{covert2020understanding} explicitly define variable importance as follows:

\begin{quote}
\textit{Feature importance should correspond to how much predictive power it provides to the model. We can then define “important” features as those whose absence degrades $m$’s performance.}
\end{quote}
In this section, we introduce a \emph{minimal axiom}, which formalizes this intuitive and widely accepted notion. We propose it as a minimal requirement for any variable importance measure (VIM). This axiom captures the idea that a feature should be considered important only if it provides unique predictive information—i.e., if its absence degrades the model’s performance.

\begin{axiom}[Minimal Axiom]
$\psi(j, P) = 0$ if and only if $X^j \indep Y \mid X^{-j}$.
\label{ax:minimal}
\end{axiom}

This axiom aligns with the goal of identifying features that are conditionally informative for prediction. Under standard conditions, it relates to the functional dependence axiom (\ref{ax:func_dep}). For instance, when performing regression, it is standard to assume an additive, independent, centered noise:

\renewcommand{\theassumption}{1\alph{assumption}}
\setcounter{assumption}{0}

\begin{assumption}[Additive noise] \label{ass:additive_noise}
$Y = m(X) + \epsilon$ with $\epsilon \indep X$ and $\mathbb{E}[\epsilon] = 0$, for some $m \in \mathcal{F}$.
\end{assumption}

Note that $m$ is indeed the conditional expectation function, since $\mathbb{E}[Y \mid X] = \mathbb{E}[m(X) + \epsilon \mid X] = m(X)$.

Similarly, in classification, we assume a model on the probability of belonging to the class: 

\begin{assumption}[Classification] \label{ass:classif}
$\mathbb{P}(Y=1\mid X)=\sigma(m(X))$ for some $m \in \mathcal{F}$ and $\sigma:\mathbb{R}\to[0,1]$ bijective.
\end{assumption}
\setcounter{assumption}{1}
\renewcommand{\theassumption}{\arabic{assumption}}

For example, taking the logistic function as $\sigma$ shows that it generalizes the logistic regression model.

Under either of these assumptions, the conditional dependence in the minimal axiom is related with functional dependence: 

\begin{propo}[Conditional and Functional Independence]\label{prop:CIT_functional_dep}
Under Assumption~\ref{ass:additive_noise} or \ref{ass:classif}, $X^j \indep Y \mid X^{-j}$ if and only if $m(X)$ is not a function of $X^j$.
\end{propo}


These assumptions are necessary to establish the equivalence. Indeed, conditional independence always implies functional independence, but the converse does not hold. For instance, in regression, if Assumption~\ref{ass:additive_noise} is violated, functional independence does not imply conditional independence. This is demonstrated by a counterexample from \cite{Ewald_2024}, where $Y \sim \mathcal{N}(X^1, X^2)$ and $X^1 \perp\!\!\!\perp X^2$. In this case, $\mathbb{E}[Y \mid X] = X^1$, implying that $Y$ is functionally independent of $X^2$. However, $Y \notindep X^2 \mid X^1$, so conditional independence does not hold. Nevertheless, Assumptions~\ref{ass:additive_noise} and \ref{ass:classif} are both standard and sufficiently general, and any statistical selection procedure involving models implicitly relies on these assumptions since they can in general only capture first-order moments. In what follows, we assume that the equivalence in Proposition \ref{prop:CIT_functional_dep} holds.

This minimal axiom is more general and helps bridge the gap between variable importance and variable selection, two concepts that are often treated separately in the literature. Variable importance methods typically distribute predictive contribution across features and then identify the top-ranked variables as important, often without formal statistical guarantees. In contrast, variable selection aims to identify a minimal subset of predictive features \citep{scornet2022mda}.

By unifying these perspectives under a common statistical framework, the minimal axiom enables both fields to benefit from one another. On the one hand, \textit{controlled} variable selection procedures provide rigorous statistical guarantees, such as control of the Type-I error rate or the False Discovery Rate \citep{candes2018, hrt}. On the other hand, variable importance measures offer richer information than a binary inclusion/exclusion decision by quantifying the relative contribution of each variable. Both perspectives are essential for scientific understanding, as they support interpretable and statistically principled conclusions regarding the roles of different variables.

Importantly, such a connection would not have been possible under the Functional Dependence Axiom (Axiom~\ref{ax:func_dep}). To the best of our knowledge, there are currently no statistical procedures specifically designed to test the corresponding null hypothesis. In contrast, a rapidly growing literature has developed powerful methods for testing conditional independence \citep{candes2018,Shah_2020, dcrt, pmlr-v206-shaer23a, Grünwald02042024}. Therefore, under the additional standard learning assumptions introduced above, we establish the equivalence between functional independence and conditional independence. This equivalence allows us to leverage the rich body of conditional independence testing methods to assess variable relevance with formal statistical guarantees.

Importantly, this approach is flexible enough to accommodate various interpretations of variable importance, depending on the inferential or practical objective. This will be illustrated in Section \ref{sect:examples}, where we examine several widely used VIMs and assess whether they satisfy the minimal axiom. 


\section{Principled approach to define a VIM}\label{sect:pipeline}
\subsection{Pitfalls of inference-based classification of VIM}
Historically, many variable importance measures have emerged as heuristics for interpreting black-box models.
As such, they often focus on quantifying the effects of perturbations or modifications to the information provided by a specific feature.
However, since many of these measures are closely tied to particular estimation procedures, the literature has introduced several ad hoc categorizations that can obscure the core statistical objective.
For example, it is common to distinguish between VIMs that rely on a single model via perturbations and those that require refitting new models. 
This distinction reflects a genuine practical consideration: refitting a model can shift what is effectively being measured, since a newly trained model may differ from the original one in ways that go beyond the single covariate being removed.
%
For instance, \cite{molnar2025} stated that Leave-One-Covariate-Out \cite[LOCO]{lei2018} and PFI \citep{Breiman2001} cannot be compared:

\begin{quote}
    \textit{LOCO differs from the other methods [...] since most of the other methods don’t require retraining the model. However, due to retraining the model, the interpretation shifts from only interpreting that one single model to interpreting the learner and how model training reacts to changes in the features.
    }
\end{quote}

Similarly, they claimed that conditional PFI  \citep[CFI,][]{Strobl2008} and LOCO are not directly comparable:

\begin{quote}
    \textit{But since conditional PFI and LOCO work differently, they differ in their interpretations. Conditional PFI is an interpretation that only involves the model at hand. LOCO importance focuses more on the machine learning algorithm, since it involves retraining the model, and the interpretation now involves multiple models trained differently.
    }
\end{quote}

This perspective is well-founded in practice: refitting a model does change the estimation procedure and a dependence on the training algorithm that a perturbation-based method does not share. However, we argue that it can be complemented by an additional perspective: the models used are surrogates for the true data-generating process, and most variable importance measures ultimately aim to estimate some underlying population-level index, regardless of how that index is estimated.

%
For example, in \eqref{eq:tsi_perturb} in Section \ref{sec:TSI}, we show that conditional PFI coincides with LOCO asymptotically, both targeting the Total Sobol Index.
%

%
We therefore propose to first characterize the theoretical estimand each method targets, and only then examine how different estimation strategies (refitting, marginalizing, perturbing) affect its finite-sample interpretation.

\subsection{Proposed approach}
\label{sec:proposed-approach}
Our proposed method consists of three steps: first, defining the theoretical index; then, estimating it; and finally, providing statistical guarantees for the important covariates.

The starting point in this pipeline is the \textbf{theoretical index} itself, as it encodes the desired interpretation of \textit{importance}. By choosing a meaningful and interpretable population quantity, one can ensure properties such as how correlation with other variables influences a feature's importance. It is at this stage that one should explicitly decide which axioms the VIM should satisfy, acknowledging the subjectivity inherent to the notion of importance. Indeed, we do not expect a single index to suit all use cases—this aligns with the "Assuming One-Fits-All Interpretability" pitfall identified by \cite{Molnar2022}. This choice will yield an insightful ranking aligned with the specifically defined goals. Moreover,  our minimal axiom should be verified at this stage, as it represents a minimal requirement for this theoretical quantity.

Given that the data-generating process is unknown in practice, the next step is to \textbf{estimate} the theoretical index. This step can involve a variety of estimation procedures, each with distinct statistical and computational properties. Some estimators aim for nonparametric efficiency \citep{williamson2023general}, others focus on reducing variance \citep{paillard2025measuringvariableimportanceheterogeneous}, achieving robustness to model misspecification \citep{lobo2025doublerobustnessconditionalfeature}, or ensuring computational feasibility \citep{LocoVSShapley}. For example, as we will discuss in the context of the Total Sobol' Index, one may estimate the index by refitting multiple models or, alternatively, by estimating the input conditional distribution. The choice of estimation method may depend on the application—for instance, in scenarios with abundant unlabeled data, one might assume complex relationships between features and the response, but simpler relationships among features themselves, as in model-X methods \citep{candes2018}. 
Moreover, since these estimates rely on complex predictive models, they suffer not only the estimation error but also the optimization error \citep{Bottou}. It is therefore desirable to account for the variability across multiple fitted models to obtain a more reliable estimate of the underlying theoretical target. This phenomenon, where many models achieve similar predictive performance while yielding different explanations, is commonly referred to as the \emph{Rashomon effect} \citep{Fisher_rashomon, rashomon}. Recently, \cite{paillard2026aggregate} showed that, in general, it is preferable to aggregate the models' predictions before computing variable importance, rather than aggregating the importance scores computed from individual models.


Because the estimation step introduces approximation error, the final estimator may not perfectly reflect the theoretical index. Consequently, to support reliable scientific discovery, it is crucial to provide \textbf{statistical guarantees} for the selected features. Depending on the context, these guarantees might take the form of asymptotic \citep{williamson2023general, lobo2025doublerobustnessconditionalfeature} or finite-sample \citep{dcrt} type-I error control, or False Discovery Rate control when conducting multiple testing \citep{candes2018}. This final step is important for ensuring reliable discoveries. Indeed, \citet{hrt} presented real-world examples where statistically grounded selected features correlated poorly with the rankings given by heuristic importance scores.

Note that these statistical guarantees are with respect to the null hypothesis, which is common to all VIMs satisfying the Minimal Axiom. In contrast, selecting the top-$k$ ranked features is a common practice that provides no guarantees on false discoveries, and the resulting ranking depends on the chosen index. In any case, recovering a correct feature ranking is primarily an estimation property rather than a property of the theoretical importance measure. For example, it may be achieved by estimators with smaller bias or variance, and there is an active line of research on statistical guarantees for feature rankings \citep{neuhof2026rankintervalsleaderboardshierarchical}. Moreover, if the goal is not variable selection but model selection—i.e., finding the smallest subset of predictive features that yields a good model—then, although both approaches aim to identify the Markov blanket, their objectives differ. In the latter case, theoretical guarantees are not the primary goal; rather, the focus is on predictive performance. Thus, such a ranking may still provide good results for this objective.

Finally, note that this pipeline contrasts with the standard Interpretable Machine Learning literature. Typically, models are first used to capture complex mechanisms, and interpretability techniques are then applied to explain the fitted model. In our pipeline, however, the goal from the outset is Scientific Discovery: using the model to understand the underlying distribution, rather than the model itself. We therefore propose to account for model estimation bias when estimating the theoretical index, rather than incorporating it into the definition of the theoretical index itself, in contrast to previous work \cite{covert_unifying, Ewald_2024}.



\subsection{The Many Facets of the Total Sobol' Index}

Although the approach proposed above is standard in Statistics, much of the variable importance literature has shifted from it. In this section, we focus on a single theoretical target—the Total Sobol' Index—and show that it admits several equivalent reformulations. Remarkably, these reformulations would place the same quantity into different categories under existing taxonomies of variable importance measures, even though they all define exactly the same estimand. 
\begin{restatable}[Equivalent reformulations of the Total Sobol' Index]{theorem}{firsteuclid}
\label{prop:tsi_reform}
Define the Total Sobol' Index by
\[
\psi_{\mathrm{TSI}}(j):=\mathbb{E}\!\left[\operatorname{Var}(Y\mid X^{-j})\right].
\]
Then,
\begin{align}
    \psi_{\mathrm{TSI}} &= \e{\mathrm{Var}(Y\mid X^{-j})}\label{eq:tsi_var}\\
    &= \e{(m_{-j}( X^{-j})-m(X))^2}\label{eq:tsi_diff_model} \\&\notag\hfill\textcolor{gray}{variance}\\
    &=\mathbb{E}\left[\left(m_{-j}(X^{-j})- Y\right)^2\right] - \mathbb{E}\left[(m(X)- Y)^2\right]\label{eq:tsi_squared}\\&\notag\hfill\textcolor{gray}{loss/refitting}\\
    &=\mathbb{E}\left[\left(\e{m(X)\mid X^{-j}}- Y\right)^2\right] - \mathbb{E}\left[(m(X)- Y)^2\right]\label{eq:tsi_marg}\\&\notag\hfill\textcolor{gray}{marginalization}\\
    &=\frac{n_\mathrm{cal}}{n_\mathrm{cal}+1}\left(\e{\left(\frac{1}{n_\mathrm{cal}}\sum_{k=1}^{n_\mathrm{cal}}m(\widetilde{X}_k^{(j)}, X^{-j})-Y\right)^2}\right.\label{eq:tsi_scpi} \\ & \left.\quad-\e{\left(m(X)- Y\right)^2}\right)\notag\\&\notag\hfill\textcolor{gray}{perturbation/marginalization}\\
    &=\frac{1}{2}\left[\mathbb{E}\left[\left(m(\widetilde{X}^{(j)}, X^{-j})- Y\right)^2\right] - \mathbb{E}\left[(m(X)- Y)^2\right]\right]\label{eq:tsi_perturb}\\&\notag\hfill\textcolor{gray}{perturbation}\\
    &=\sigma^2(R^2-R^2_{-j}),\label{eq:tsi_func_anova}
\end{align}
where the expectation in \eqref{eq:tsi_perturb} is taken with respect to
\[
P_{(\widetilde{X}^{(j)},X^{-j},Y)}
=
P_{X^j\mid X^{-j}}
\times
P_{(X^{-j},Y)},
\]
and the expectation in \eqref{eq:tsi_scpi} is taken over
$n_{\mathrm{cal}}$ independent samples from this distribution.
\end{restatable}


Notably, none of these equalities requires the additive noise assumption (Assumption~\ref{ass:additive_noise}). From these equalities, we first note the richness of the Total Sobol' Index (TSI): it admits multiple interpretations and computational forms. This highlights the value of the general estimation pipeline—despite differences in how the index is computed, each formulation targets the same underlying quantity. Yet, the current literature often categorizes these methods into distinct, seemingly incompatible families, claiming that they are not comparable \citep[see][]{Ewald_2024, molnar2025, fumagalli2025unifyingfeaturebasedexplanationsfunctional}.

First, from \eqref{eq:tsi_var} to \eqref{eq:tsi_squared}, we observe that the distinction between \textit{variance}-based and \textit{loss}-based approaches is artificial. Although variance-based methods do not explicitly use the output $Y$ in estimation \eqref{eq:tsi_diff_model}, they nonetheless target the same quantity as loss-based methods based on \eqref{eq:tsi_squared}, because the dependence in $Y$ is hidden in the models $m$ and $m_{-j}$. This artificial distinction was also made by \cite{covert_unifying}, who framed it in terms of the aspect of model behavior being explained. Specifically, \eqref{eq:tsi_squared} corresponds to explaining the model's \textit{dataset loss}, whereas \eqref{eq:tsi_diff_model} corresponds to explaining the \textit{dataset loss with respect to the output}.

Equation \eqref{eq:tsi_squared} presents the Total Sobol' Index using the squared loss based on two different models $m$ and $m_{-j}$. Equation \eqref{eq:tsi_marg} corresponds to a marginalization-based approach, similar to that of \cite{covert2020understanding}. Also, \cite{covert_unifying} proposed it as a way of avoiding refitting. The plug-in in this formulation is often labeled a \emph{one-function} variable importance measure (VIM), since the same estimated model $m$ is used on both sides of the comparison, without retraining a separate $m_{-j}$ for each $j$, which provides an advantage in terms of computation. 

However, this perspective overlooks a critical challenge: the conditional expectation $\mathbb{E}[m(X) \mid X^{-j}]$ is generally unknown and must itself be estimated. This requires both estimating the conditional distribution and the conditional expectation. This estimation framework was formalized in the Sobol-CPI method proposed by \cite{lobo2025doublerobustnessconditionalfeature}, who also demonstrated that not having an infinite number of conditional samples introduces bias. This gives \eqref{eq:tsi_scpi} with a fix calibration set size $n_\mathrm{cal}$.

Equation \eqref{eq:tsi_perturb} falls under the class of \emph{perturbation methods}, since it uses a fixed model $m$ and perturbs the $j$-th coordinate by drawing from the conditional distribution— a normalized version of the Conditional Feature Importance \citep[CFI,][]{Strobl2008}. 
From \eqref{eq:tsi_func_anova}, we note that this index can also be interpreted as the difference in nonparametric $R^2$ \citep{williamson2021}.

Taken together, these results reveal that the usual categorization into refitting, perturbing, and marginalizing is misleading: in this example, all approaches aim to estimate the same functional. As such, they share the same interpretation and comparing them is meaningful.

The estimation and the statistical guarantees for this index will be further discussed in Section \ref{sec:TSI}.

\section{Application of the framework}\label{sect:examples}

The framework defined in Section~\ref{sect:pipeline} is intended to guide practitioners not only in selecting an index aligned with their goals, but also in choosing an estimation procedure that balances statistical robustness, computational constraints, and inferential guarantees. In this section, we provide an extensive study of \textbf{theoretical indices}, discuss their interpretability and whether they satisfy the minimal axiom, and examine different \textbf{estimation} strategies adapted to the data at hand along with their \textbf{statistical properties}. This principled approach sheds some light on the interpretation of classical indices.


In particular, we revisit the commonly held belief that Permutation Feature Importance  \citep[PFI,][]{Breiman2001} is suited only for marginal importance \citep{Ewald_2024, molnar2025}. This belief is a natural consequence of how PFI is constructed: since the permutation is performed marginally, it seems intuitive that the resulting index should only be informative about marginal, rather than conditional, independence. We show, however, that this intuition does not hold for the theoretical index PFI targets: instead, PFI aligns with conditional importance, and therefore satisfies the minimal axiom. This arises because an implicit conditional filtering emerges during the training of the model, which we detail below.

This discrepancy between the marginal construction of PFI and the conditional nature of its theoretical target stems from a lack of clarity, in the existing literature, about the null hypothesis underlying PFI.
It is also caused by terminology, especially since a conditional counterpart exists \citep{Strobl2008,Hooker2021}. 
Similar terminology-driven distinctions are discussed below, where we show that the marginal versions of SAGE are in fact suitable for conditional importance, while the conditional SAGE, which coincides with the standard Shapley values, is not.

All the proofs can be found in Appendix \ref{sect:ma_proofs}.

\subsection{Coefficients of a generalized linear model}

The first, most intuitive importance index relies on the coefficients of a linear model. It is simple to interpret because the effect of each variable can be directly inferred from the magnitude of its coefficient. Moreover, the same reasoning applies to logistic regression and, more broadly, to generalized linear models (GLMs). These models are often favored for their interpretability \citep{Rudin2019,Molnar2022} and have been advocated over black-box models, especially in sensitive applications such as hate speech detection \citep{reyero2023knowledge}. In this section, we argue that this simple variable importance measure, along with some modifications, satisfies the minimal axiom.

We begin by stating the standard generalized linear model (GLM) assumption:

\begin{assumption}[GLM assumption]\label{ass:glm}
Given a link function $g$, we assume
\[
g\left(\mathbb{E}[Y \mid X]\right) = X^1 \beta_1 + \ldots + X^p \beta_p.
\]
\end{assumption}

Under this assumption, as introduced in \cite{mccullagh1989generalized}, a natural variable importance measure for covariate $j$ is given by the magnitude of the coefficient. Thus, the \textbf{theoretical index} is given by: 

\begin{definition}[GLM indices] Given $j\in [p], P\in \mathcal{M}$, $\psi_\mathrm{GLM}(j, P)$ is defined as
\begin{align*}
    \psi_\mathrm{GLM}(j):=\beta_j^2.
\end{align*}
\end{definition}

The square is used to ignore whether the effect is positive or negative; what matters is simply that there is an effect. Also, we implicitly assume that the input features have been standardized. 
This measure, clearly aligned with Axiom \ref{ax:lin_agree}, can be useful in some settings due to its simplicity and the direct interpretability of the coefficients, which represent the contribution of each covariate to the predictive function.
This simple VIM satisfies the minimal axiom:

\begin{propo}\label{prop:glm}
Under the GLM assumption (\ref{ass:glm}), $\psi_{\mathrm{GLM}}$ satisfies the minimal axiom.
\end{propo}


However, even if this minimal axiom is fulfilled for this importance index, in practice there may be \textbf{estimation} complications due to high dimensionality and/or collinearity between input variables. These are issues of inference rather than interpretation problems with the theoretical index. For this reason, many penalization strategies have been introduced, such as the Akaike Information Criterion \citep[AIC,][]{AIC} or cross-validation controlled $\ell_1$ penalization used in Lasso \citep{lasso}. 
%
We observe that all these strategies still satisfy the minimal axiom, as the theoretical index continues to assign zero importance to unimportant features. Moreover, there are numerous results on bounding the number of false discoveries made by the inference procedures used in these methods (see, for example, Corollary 5.3 from \cite{Giraud2021}, which provide bounds in probability).

Many variable selection procedures with their respective \textbf{statistical guarantees} can be derived from this variable importance index. For instance, to control the type-I error using Conditional Randomization Tests  \citep[CRT,][]{candes2018}, the most effective procedures are based on the difference of coefficients from a Lasso model trained on the original data versus conditional independent data \citep{dcrt}. One can easily observe that the sum over the conditional samplings tends to the Lasso index, as the coefficients of the conditional independent generated input coordinates will tend to zero. Therefore, the minimal axiom of the given theoretical index is satisfied with finite sample guarantees on the selected features. Similarly, this approach can be applied to the Knockoffs framework with the popular Lasso Coefficient Difference \citep[LCD,][]{candes2018}, which provides False Discovery Rate control on the selected set. Finally, additional guarantees on $p$-values or on the familywise error rate can be obtained from corrected versions of the Lasso or Ridge \citep{ridge_corr, Dezeure2015}.

\subsection{Some popular perturbation indices (constant/marginal/conditional imputation)}\label{sect:perturb}

In the general case, assuming a linear model is not sufficient, and we may want to adapt to more complex settings. Moreover, if the model does not generalize well, there may be no reliable scientific inference, as the model acts merely as a surrogate for the data-generating process.
Thus, we are unable to provide interesting insights on feature importance with a model that does not represent the underlying distribution. This relates to the poor model generalization pitfall described by \cite{Molnar2022}. Therefore, there is a need for model-agnostic theoretical indices. In the sequel, we present model-agnostic VIMs that use the loss to measure the predictability of the model when incorporating the information of the $j$-th covariate versus when it is not used. Consequently, a data-splitting step is required, which may decrease the power of the procedure \citep{dcrt}.

The first approach we present involves reusing the same model while perturbing the inputs to observe the effect of each coordinate on the model’s output. Specifically, the model's performance with the $j$-th coordinate perturbed is compared to its performance with the original input. They initially appear to fall outside the proposed principled approach. Indeed, they are often introduced as heuristics for model interpretation. Thus, they are typically defined as theoretical quantities that depend on a specific model (\cite{covert_unifying, Ewald_2024}), rather than as theoretical indices for which the model is part of the estimation procedure. Consequently, the interpretation of the estimated quantity can depend on the bias of the specific model, making the corresponding null hypothesis unclear. For instance, although Permutation Feature Importance was introduced in \cite{Breiman2001}, it was not until \cite{scornet2022mda} that its estimand was formally studied. To address this issue, we proceed within the proposed principled framework.


We begin by defining the \textbf{theoretical index}:

\begin{definition}[Perturbation indices]\label{def:perturb} Given $j\in [p], P\in \mathcal{M}$, $\psi_\mathrm{perturb}(j, P)$ is defined as
\begin{align*}
    \psi_\mathrm{perturb}(j):=\e{\ell(Y, m(\check{X}^{(j)}, X^{-j}))}-\e{\ell(Y, m(X))},
\end{align*}
where $\check{X}^{(j)}$ is a \textit{perturbed} version of $X^j$
and therefore the joint distribution changes, i.e., $P_{(X^j, X^{-j}, Y)} \neq P_{(\check{X}^{(j)}, X^{-j}, Y)}$.
\end{definition}

All the coordinates are preserved except for the $j$-th coordinate, which is sampled in a way that alters the joint distribution, making it different from the original one. Each perturbation corresponds to a different index.

Now, we present several examples of perturbations along with their corresponding \textbf{estimation} procedures. 

The most popular perturbation-based index is the permutation feature importance  \citep[PFI,][]{Breiman2001, Mi2021} due to its simple estimation. This approach relies on a perturbation denoted by $X^{(j)}$ such that
\[
P_{(X^{(j)}, X^{-j}, Y)} = P_{X^j} \times P_{(X^{-j}, Y)}.
\]
Then, $X^{(j)}$ preserves the marginal distribution ($P_{X^{(j)}} = P_{X^j}$), and it is independent of the output ($X^{(j)} \indep Y$) as well as of the remaining input variables ($X^{(j)} \indep X^{-j}$). The theoretical index associated with PFI was studied by \cite{scornet2022mda}, who showed that it is not always desirable, as it can lead to misleading conclusions in the presence of correlation. This reinforces our proposal to study the theoretical index carefully, since otherwise we risk misunderstanding what is considered important.

Regarding its estimation, the simplest method consists of permuting the value across observations. However, this approach also suffers from inference issues because it induces extrapolation \citep{Hooker2021}. Indeed, due to permutation, the model attempts to predict in low-density regions where it was not trained, resulting in unpredictable behavior. This estimating issue also happens with the constant perturbation and, therefore, directly predicting on an input in which the $j$-th feature has been replaced by a constant $a \in \mathbb{R}$, for instance $0$ \citep{cxplain}.

To address this estimation issue, \cite{Hooker2021, chamma2023statistically} proposed permuting the $j$-th coordinate conditionally to the other ones, leading to the Conditional Permutation Importance (CPI). This method was primarily introduced to tackle the inferential problem, without thoroughly studying the theoretical quantity to which it converges. This theoretical aspect was analyzed in \cite{lobo2025doublerobustnessconditionalfeature}, where it was shown that the conditional permutation step corresponds to sampling from the conditional distribution; then, CPI corresponds to Conditional Feature Importance \cite[CFI,][]{Strobl2008}). We denote this conditional perturbation by $\widetilde{X}^{(j)}$. It is given by $$P_{(\widetilde{X}^{(j)}, X^{-j}, Y)}=P_{X^j\mid X^{-j}}\times P_{(X^{-j}, Y)}.$$
Therefore, CFI assumes that the $j$-th coordinate is independent of the output given the other inputs, i.e., $\widetilde{X}^{(j)} \indep y \mid X^{-j}$, and preserves the conditional distribution, $P_{(X^j \mid X^{-j})} = P_{(\widetilde{X}^{(j)} \mid X^{-j})}$. Moreover, \cite{lobo2025doublerobustnessconditionalfeature} showed that for the quadratic loss, CFI coincides up to a explicit universal constant with the Total Sobol' Index, presented below. Also, they showed that it is possible to sample from this conditional distribution by first regressing $X^j$ on $X^{-j}$ and then adding a random residual under what they called the additive innovation, which consists in assuming homoskedastic noise for each input with respect to the rest. Estimating these conditional distributions is an active area of research for their application in the Model-X framework, and there are several options, such as domain-specific samplers \citep{sesia2019gene} or others based on deep neural networks \citep{romano2020deep, zhu2021deeplink}.

Other types of perturbations can also be considered, such as relational perturbations \citep{RFI}, which operate similarly to the CFI but condition on a smaller subset of coordinates. However, each perturbation gives rise to a different theoretical index, and thus it is not meaningful to compare them directly, as they pursue different objectives and consequently yield different rankings of importance. 

In any case, these indices satisfy the minimal axiom for the previous presented perturbation. 
In addition, under the following assumption from \citet{lobo2025doublerobustnessconditionalfeature}, the estimated index is also able to correctly identify the null coordinates.


\begin{assumption}[Asymptotic relevance]\label{ass:asymp_relevance}
Denote by $\check{X}^{(j)}$ a random variable sampled from a distribution $\check{P}_j$. Then there exists a sequence $(\delta_i)_{i\in \mathbb{N}}$ and an $n_0\in \mathbb{N}$ such that for any $n\geq n_0$ we have that
\[
\mathbb{E}\!\left[\left|\hat{m}_n\!\left(\check{X}^{(j)}, X^{-j}\right)-\hat{m}_n(X)\right|\right]
\leq \delta_n,
\]
where $\delta_n \to 0$ as $n \to \infty$.
\end{assumption}

We observe that this assumption explicitly requires the model to be independent of the null coordinates, and it can be obtained under standard assumptions on the model. In particular, there is no extrapolation under the null hypothesis \citep[see][]{lobo2025doublerobustnessconditionalfeature}. We further discuss this assumption in Appendix~\ref{sect:asymp_rel}, where we explore the functional dependence on the null features of a large range of standard machine learning models, such as Gradient Boosting, Random Forests, and Neural Networks, including more complex models such as the SuperLearner \citep{van2007super} and the recent state-of-the-art prediction model for tabular data, TabICL \citep{qu2026tabiclv2}, across many settings for regression and classification and under stronger correlation regimes.

\begin{propo}[$\psi_{\mathrm{perturb}}$ satisfies the minimal axiom]\label{prop:perturb_ma}
Under additive noise assumption (\ref{ass:additive_noise}) or classification assumption (\ref{ass:classif}), $\psi_{\mathrm{perturb}}$ satisfies the minimal axiom for any strictly convex loss $\ell$ using constant, marginal or conditional perturbation. Moreover, under asymptotic relevance assumption (\ref{ass:asymp_relevance}), if $X^j \indep Y \mid X^{-j}$, then the plug-in estimate satisfies $\hat{\psi}_{\mathrm{perturb}} \to 0$ almost surely for any bounded continuous loss $\ell$ and continuous model.
\end{propo}
%
%

When applied to PFI, the first part of the proposition implies, that PFI is also suitable for detecting conditional independence. 

This result contrasts with the existing literature: since the perturbation in PFI is marginal—also with respect to the rest of the input—it is often claimed that PFI is only suited for detecting marginal independence, not conditional independence \cite[see][]{Ewald_2024, molnar2025}. 
%
In general, \cite{Ewald_2024} first showed that if a method is suited for marginal independence, then it cannot be suited for conditional independence. Then, under strong assumptions, they related PFI to marginal independence and therefore claimed that it could not be used for conditional independence. 
In Appendix~\ref{subsect:PFI_marginal}, we prove that the assumption made in \cite{Ewald_2024} to draw conclusions about marginal independence is strong enough to imply conditional independence as well. We also provide counterexamples in which PFI is zero (due to conditional independence) despite marginal dependence, as well as examples in which PFI is strictly positive (due to conditional dependence) despite marginal independence.


Note that, for any perturbation, we have that under conditional independence the Bayes predictor does not depend on the coordinate, since it is not predictive, and therefore the index is zero. For the conditional dependence case, we need to make assumptions about the specific perturbation.
Indeed, there are perturbations for which the model is invariant yet still depends on the feature. For instance, for a model of the form $m(X)=X^j(1-X^j)$, it is invariant under the perturbation $\widetilde{X}^j := 1 - X^j$. Therefore, to ensure that the index is non-zero under conditional dependence, we need to specify the particular perturbation.

It is important to note that this key result comes from a form of equivalence between functional and conditional independence (see Proposition~\ref{prop:CIT_functional_dep}). Consequently, the trained model $\widehat{m}$ is not just any model; it is a loss minimizer that approximates the data-generating process and thus serves as an initial filter for conditional independence.

Regarding the second part of Proposition~\ref{prop:perturb_ma}, for inference purposes, when using a consistent estimator, the estimated importance of irrelevant coordinates vanishes asymptotically. This follows from Theorem 3.1 in \cite{lobo2025doublerobustnessconditionalfeature}, which also states that when using CFI—that is, when the perturbation is conditional—there is a double robustness property. Specifically, it suffices for only one of the models involved to be consistently estimated in order to detect the null hypothesis, which provides a valuable inference guarantee. 

Finally, regarding the \textbf{statistical properties} of such methods, it is important to note that satisfying the minimal axiom does not provide any finite-sample guarantee, since this property relates exclusively to the theoretical index. Therefore, even though the PFI satisfies the minimal axiom, to the best of our knowledge there is no fully valid available test. Indeed, \cite{chamma2023statistically} showed that directly using a $t$-test does not control the type-I error, particularly under high correlation. This is due to the vanishing influence function under the null, which prevents a Central Limit Theorem, similarly to what happens for other quadratic functionals \citep{williamson2021, LocoVSShapley}.

Some asymptotic results for the CFI regarding type-I error have been established by \cite{lobo2025doublerobustnessconditionalfeature}, based on Markov's inequality using a variance correction. Moreover, similarly to the case of Lasso coefficients, certain CRT-based methods can be adapted to provide finite-sample type-I error control, for example by using the Holdout Randomization Test from \cite{hrt}. Following the ideas of \citet{Watson2021}, other nonparametric tests such as the sign test or the Wilcoxon test can also be considered to obtain finite-sample type-I error control. Finally, \cite{lobo2026semiknockoffs} proposed Semi-knockoffs, which also provide finite-sample type-I error and FDR control by comparing two distributions: one perturbed using only the information from the input, and another using both the input and the output.


\subsection{Generalized Total Sobol' Index}\label{sec:TSI}

The Total Sobol' Index (TSI) was originally introduced in the context of sensitivity analysis by \cite{first-total-order-effects} as the proportion of output variance attributed to a specific input coordinate when the remaining inputs are known. This notion can be extended beyond variance to accommodate any loss function $\ell$ \citep{williamson2023general}. The \textbf{theoretical index} is given by:

\begin{definition}[Generalized Total Sobol' Index (TSI)]\label{def:TSI} Given $j\in [p], P\in \mathcal{M}$ and a loss function $\ell$,  the Generalized Total Sobol' Index is defined as
\begin{align*}
    \psi_{\mathrm{TSI}}(j) := \mathbb{E}\left[\ell\left(m_{-j}(X^{-j}), Y\right)\right] - \mathbb{E}\left[\ell(m(X), Y)\right],
\end{align*}
where $m_{-j}(X^{-j}):=\e{Y\mid X^{-j}}$ and $m(X):=\e{Y\mid X}$.
\end{definition}

This index is a widely used measure of variable importance \cite[see, e.g.,][]{lei2018, Rinaldo2019, Hooker2021, scornet2022mda, williamson2023general}. It is termed \emph{generalized} because, when the loss function $\ell$ is the squared error, the expression recovers the unnormalized TSI definition. 

Notably, with the quadratic loss, TSI can also be rewritten in several equivalent forms as seen in Proposition \ref{prop:tsi_reform}. Their differences lie in inference properties rather than conceptual targets. We observe that it can be \textbf{estimated} as a direct plug-in using any of these equivalent formulations. For example, \cite{williamson2021} studied the difference between plug-in estimators used in \eqref{eq:tsi_diff_model} and \eqref{eq:tsi_squared}, showing that the former requires a one-step correction to achieve nonparametric efficiency, while the latter does not. The method in \eqref{eq:tsi_squared} corresponds to Leave-One-Covariate-Out (LOCO). Furthermore, the Permute-and-Relearn and Condition-and-Relearn importances of \citet{Hooker2021}, which consist of either marginally or conditionally permuting a feature before relearning the model, also fall within this category, since they aim to estimate 
$$\e{Y\mid X^{-j},\widetilde{X}^j}=\e{Y\mid X^{-j}}=m_{-j}(X^{-j}).$$

If a plug-in method is applied to \eqref{eq:tsi_marg} or equivalently \eqref{eq:tsi_scpi}, the resulting method is known as Sobol-CPI($n_\mathrm{cal}$) \citep{lobo2025doublerobustnessconditionalfeature}, which was shown to be nonparametrically efficient. It also corresponds to the conditional SAGE value function \citep{covert2020understanding}. Using the estimator in \eqref{eq:tsi_perturb} leads to $0.5\times$CFI \citep{Strobl2008}. It is also known as Sobol-CPI(1) \citep{lobo2025doublerobustnessconditionalfeature},  and they proved that this estimator is \emph{double robust}, meaning that it can reliably identify null covariates as long as either the predictive model or the conditional sampler is well-specified—thereby reducing the risk of false positives and making it a strong candidate for scientific discovery.  

Another insightful interpretation of the Total Sobol' Index arises when using the cross-entropy loss. In this case, it becomes directly connected to information-theoretic quantities such as mutual information (denoted by $\mathrm{I}$) and the Kullback–Leibler (KL) divergence, as discussed by \cite{covert2020understanding}. Specifically, we have
\begin{align}\label{eq:TSI_CE}
    \psi_{\mathrm{TSI}}(j) &= \mathrm{I}(Y; X^j \mid X^{-j}) \\&= D_{\mathrm{KL}}\left(P_{(Y, X^j \mid X^{-j})} \,\big\|\, P_{(Y \mid X^{-j})} P_{(X^j \mid X^{-j})}\right)\notag,
\end{align}

which quantifies the conditional mutual information between $Y$ and $X^j$ given $X^{-j}$. Intuitively, this measures the reduction in uncertainty about $Y$ obtained by adding $X^j$ to the already known covariates $X^{-j}$.

Finally, we note that the Total Sobol' Index satisfies the minimal axiom.
\begin{propo}\label{prop:TSI_ma}
Let $\ell$ be a loss with a unique minimizer which is a function of the conditional distribution of $Y$ given $X$. Under additive noise assumption (\ref{ass:additive_noise}) or classification assumption (\ref{ass:classif}), then, $\psi_{\mathrm{TSI}}(j, P)$ satisfies the minimal axiom.
\end{propo}

As discussed in \cite{williamson2023general}, most commonly used losses fulfill the condition stated in Proposition \ref{prop:TSI_ma}, with their Bayes-optimal predictors being functions of the conditional expectation. Examples include the mean squared error (MSE), deviance, classification accuracy, and the area under the ROC curve (AUC).

This makes the Total Sobol' Index a suitable criterion for variable selection, as also supported in the literature: \cite{scornet2022mda} stated that it is the best quantity for finding the minimal subset of predictive variables.

Leveraging the relationship with CFI, it is possible to directly inherit all the \textbf{statistical guarantees} established in the previous section, such as the asymptotic Type-I error control based on Markov’s inequality \citep{lobo2025doublerobustnessconditionalfeature} and the finite-sample guarantees from the HRT \citep{hrt}. Furthermore, when analyzing the LOCO version, one can employ the data-splitting variants proposed by \cite{williamson2023general} to achieve asymptotic Type-I error control or, similarly to the CFI, rely on the Markov-based guarantees \citep{LocoVSShapley}. Finally, similarly to the perturbation indices, nonparametric tests such as the sign test or the Wilcoxon test can be directly applied by treating the losses using the $j$-th coordinate as one population and those excluding this information as another. In this way, we obtain finite-sample type-I error guarantees \citep{lei2018,Watson2021}.

\subsection{Modifications of Total Sobol' Index}
\cite{LocoVSShapley} criticized the TSI due to its tendency to underestimate the importance of features that are correlated with others. This phenomenon is known as \emph{correlation distortion}, see \cite{dLOCO}. To mitigate this issue, \cite{LocoVSShapley} proposed a normalized \textit{decorrelated} version of the TSI, defined as
\[
\psi_{\mathrm{dTSI}}(j) := \frac{\mathbb{E}\left[(m(X) - m_{-j}(X^{-j}))^2\right]}{\mathbb{E}\left[(X^j - \nu_{-j}(X^{-j}))^2\right]},
\]
where $\nu_{-j}(X^{-j}) := \mathbb{E}[X^j \mid X^{-j}]$ denotes the conditional mean of $X^j$ given all other covariates.

The motivation behind this normalization comes from the behavior of TSI under a linear model (\ref{ass:glm}). In that case, the TSI can be expressed as
\[
\psi_{\mathrm{TSI}}(j) = \beta_j^2 \, \mathbb{E}\left[(X^j - \nu_{-j}(X^{-j}))^2\right],
\]
which shows that the index is scaled by the conditional variance of $X^j$ given $X^{-j}$. Therefore, when this variance is small due to high correlation, the importance decreases. The proposed normalization corrects this by dividing out the conditional variance, thereby restoring the importance of such variables and better aligning with the axioms discussed in section \ref{subsec:verd_axioms}.

\cite{LocoVSShapley} also introduced other decorrelation-based strategies to address correlation distortion. However, they also found those alternatives to suffer from second-order bias and low-density inference issues. In any case, they satisfy the minimal axiom due to its alignment with the functional dependence axiom (see Axiom \ref{ax:func_dep} and Proposition \ref{prop:CIT_functional_dep}).

Finally, \cite{du2025disentangledfeatureimportance} proposed Disentangled Feature Importance (DFI), which is mainly based on the total Sobol' index computed in a latent space where input coordinates are rendered independent via a transport map. However, DFI only satisfies the minimal axiom in this transformed space—not in the original input space. As a result, it can inflate the importance of irrelevant coordinates and lead to a high rate of false discoveries, as seen in their experiments.

\subsection{Shapley Additive Global importancE (SAGE)}\label{subsec:Shap_not_minimal}
Shapley values are widely used for local variable importance \citep{shap, spadaccini2025hypothesisfreediscoveryepidemiologicaldata}, and were adapted to global variable importance by \cite{covert2020understanding}, who explicitly defined variable importance in terms of the predictive power a feature provides to the model.
However, in this section, we argue that this performance criterion is in fact not fulfilled by standard Shapley values, as they do not satisfy the \emph{minimal} axiom. We begin by explicitly introducing SAGE:

\begin{definition}[$\psi_\mathrm{SAGE}$]
Shapley Additive Global importancE (SAGE) is given by 
\begin{align}
    \psi_\mathrm{SAGE}(j):=\sum_{S\subset -\{j\}}w_S(v(S\cup \{j\})-v(S)),
\end{align}
where $v(S):=\e{\ell(Y, \e{Y})}-\e{\ell(Y,m_S(X^S))}$, with $m_S(X^S):=\e{Y\mid X^S}$ and $w_S:=\frac{1}{p}\binom{p-1}{|S|}^{-1}$.
\end{definition}

Each $v(S)$ denotes the SAGE value function associated with the subset $S \subset [p]$, i.e., the change in performance relative to the average prediction when incorporating the information from $X^S$. In particular, we denote by $\psi_{\mathrm{SAGEvf}}$ the value corresponding to $S = \{j\}$.

\begin{definition}[$\psi_\mathrm{SAGEvf}$]
SAGE value function associated with the $j$-th feature is given by
\begin{align*}
    \psi_\mathrm{SAGEvf}:=v(\{j\})=\e{\ell(Y, \e{Y})}-\e{\ell(Y,m_j(X^j))}.
\end{align*}
\end{definition}

Notably, it coincides with the index estimated by the Leave One Covariate In (LOCI) method. 
However, these two approaches were not previously connected—similarly to the case of LOCO and Sobol-CPI—because SAGE is based on marginalization, whereas LOCI relies on refitting. Despite this inference difference, both aim to estimate the same underlying quantity. As a result, they will often lead to the same covariate rankings and selections.

Nonetheless, we emphasize that neither the SAGE value function (and thus LOCI; see section \ref{sect:LOCI}) nor SAGE itself satisfies the minimal axiom.
\begin{propo}\label{prop:SAGE}
Neither $\psi_\mathrm{SAGE}$ nor $\psi_\mathrm{SAGEvf}$ satisfies the minimal axiom. 
\end{propo}
%
In particular, Shapley values may assign importance to features that are not actually used by the model, simply because they are correlated with truly important features (see Figure \ref{fig:motivating_ma}). Consequently, the removal of such features does not degrade model performance, since they do not directly influence the predictions of the model. Moreover,  the explainability of the method is hindered by the fact most features are assigned non-null importance: the explanations lack parsimony, and the set of features called important becomes unmanageable. 


\subsubsection{Surplus SAGE:}

\citet{Ewald_2024} also proposed to study another quantity, namely the \textit{surplus}, defined as the difference between the SAGE value function of all coordinates except the $j$-th and that of all coordinates.

\begin{definition}[$\psi_\mathrm{scSAGE}$]
Surplus SAGE value function associated with the $j$-th feature is given by
\begin{align*}
    \psi_{scSAGE}(j):=v(-j\cup j)-v(-j)=v([p])-v(-j).
\end{align*}
\end{definition}

We note that this VIM does not introduce a new index, it exactly coincides with the Total Sobol' Index (expanding the definitions, we recover \eqref{eq:tsi_marg}). Thus, we do not restate the principled approach, as it has already been applied in Section~\ref{sec:TSI}. In particular, this VIM satisfies the minimal axiom.


\begin{propo}\label{prop:scSAGEj}
Under additive noise assumption (\ref{ass:additive_noise}) or classification assumption (\ref{ass:classif}), a strictly convex loss $\ell$, $\psi_\mathrm{scSAGE}(j)$ satisfies the minimal axiom. 
\end{propo}

\subsubsection{Marginal extensions:}
in any case, the inference of $\psi_\mathrm{SAGE}$ is complex due to two main challenges:  
1) the number of subsets to consider grows exponentially with the number of features, and  
2) for each subset $S \subset [p]$, it is necessary to estimate the restricted model $m_S$.

To address the first issue, various approaches have been proposed, ranging from Monte Carlo \cite{covert2020understanding} to more sophisticated methods such as importance sampling guided by prior information, for example from the structure of random forests \cite{shaff}.

Regarding the second issue, \cite{covert2020understanding} proposed using marginalization instead of refitting. That is, rather than training a new model $\widehat{m}_S$, they approximate $m_S$ by the conditional expectation $\mathbb{E}[m(X) \mid X^S]$. For the original SAGE estimator, this requires conditional sampling, which is computationally expensive. To mitigate this, they further proposed using marginal sampling as a proxy for conditional sampling. Accordingly, they distinguish between two variants: the original \emph{conditional SAGE}, and the more computationally efficient \emph{marginal SAGE} (with similar distinctions for their value function counterparts).

However, these variants target different theoretical quantities and thus have different interpretations. In particular, while conditional SAGE and conditional SAGEvf do not satisfy the minimal axiom (as illustrated by Figure \ref{fig:motivating_ma}), marginal SAGE and marginal SAGEvf do.

\begin{propo}\label{prop:mSAGE} Let $\ell$ be a loss with a unique minimizer which is a function of the conditional distribution of $Y$ given $X$. Under additive noise assumption (\ref{ass:additive_noise}) or classification assumption (\ref{ass:classif})
$\psi_\mathrm{mSAGE}$ and $\psi_\mathrm{mSAGEvf}$ satisfy the minimal axiom.
\end{propo}

Similarly to the marginal interpretation of PFI discussed in Section \ref{sect:perturb}, \cite{Ewald_2024} related the marginal SAGE counterpart to marginal independence under strong assumptions. However, Proposition \ref{prop:mSAGE} shows that this interpretation is not accurate. In fact, similarly to the perturbation indices, the global model acts as a first filter for conditional independence through its functional dependence on the input features (see Proposition~\ref{prop:CIT_functional_dep}). Indeed, \cite{covert2020understanding} already observed that functional independence implies $\psi_\mathrm{mSAGE}=0$, but the framework presented here allows us to relate this property to conditional independence.

Intuitively, since marginal SAGE involves comparing subsets by altering only coordinate $j$, if feature $j$ is not important, it will not influence the model's output. As a result, all comparisons involving $j$ yield zero importance. 

This confusion between marginal SAGE and marginal independence partly arises from the terminology “marginal” and “conditional” SAGE. Conversely, the conditional SAGE value function, which coincides with LOCI, can only be employed for marginal testing. This highlights that the distinction is not merely theoretical but also reflects conceptual differences, as the two approaches assess feature importance either conditionally or marginally, even though both involve marginalization. Consequently, \textit{directly comparing mSAGE and cSAGE is not meaningful}.

Finally, note that, as shown in the second part of Proposition~\ref{prop:perturb_ma}, under the asymptotic relevance assumption (\ref{ass:asymp_relevance}), perturbation indices assign vanishing importance to the null coordinates. The same result can therefore be obtained under this assumption for both $\widehat{\psi}_\mathrm{mSAGE}$ and $\widehat{\psi}_\mathrm{mSAGEvf}$. 

\section{Summary of theoretical indices and VIM methods}\label{sect:overview}

In this section, we present a summary overview of the definitions of the main theoretical indices, whether they satisfy the minimal axiom, and the main Variable Importance Measures (VIMs), detailing what they estimate and how they are estimated.


We denote the conditional and marginal Shapley value functions for $S\subset[p]$ as:
\begin{align*}
v(S) &:= \mathbb{E}\left[\ell(Y, \mathbb{E}[Y])\right] - \mathbb{E}\left[\ell\left(Y, \mathbb{E}[m(X) \mid X^S]\right)\right],\\ \quad
v^m(S) &:= \mathbb{E}\left[\ell(Y, \mathbb{E}[Y])\right] - \mathbb{E}\left[\ell\left(Y, \mathbb{E}[m(X^{(-S)}) \mid X^S]\right)\right].
\end{align*}
The former ($v(S)$) reflects the intrinsic conditional dependence, while the latter ($v^m(S)$) involves marginal dependence. Therefore, $\mathbb{E}[m(X^{(-S)}) \mid X^S]$ means that the $-S$ coordinates are \textit{averaged} without taking into account the relationship with $X^S$, which are fixed. We also recall the Shapley weighting:
\[
w_S := \frac{1}{p} \binom{p-1}{|S|}^{-1}.
\]

The main theoretical indices, along with their definitions and whether or not they satisfy the minimal axiom (MA, Axiom \ref{ax:minimal}), are summarized in table \ref{tab:th_ind}.
\begin{table*}[t]
    \centering
    \caption{\textbf{Summary of the theoretical indices:} Theoretical indices, their target quantities, and whether they satisfy the minimal axiom.}
  \begin{tabular}{ccc}
\hline
\textbf{Index} & \textbf{Definition} & \textbf{MA}  \\
\hline
$\psi_\mathrm{TSI}$ & $\e{\ell(m_{-j}(X^{-j}),Y)}-\e{\ell(m(X), Y)}$ & Yes \\
\hline
$\psi_\mathrm{SAGE}$ & $\sum_{S\subset -\{j\}}w_S\left(\e{\ell(Y,\e{m(X)\mid X^S})}-\e{\ell(Y,\e{m(X)\mid X^{S\cup\{j\}}})}\right)$ & No \\
\hline
$\psi_\mathrm{LOCI}$ & $\e{\ell(Y, \e{Y})}-\e{\ell(Y,m_{j}(X^j))}$ & No \\
\hline
$\psi_\mathrm{mSAGEvf}$ & $\e{\ell(Y, \e{Y})}-\e{\ell(Y,\e{m(X^{(-j)})})}$ & Yes  \\
\hline
$\psi_\mathrm{mSAGE}$ & $\sum_{S\subset -\{j\}}w_S\left(\e{\ell(Y,\e{m(X^{(-S)})\mid X^S})}-\e{\ell(Y,\e{m(X^{(-S)})\mid X^{S\cup\{j\}}})}\right)$ & Yes  \\
\hline
$\psi_\mathrm{PFI}$ & $\e{\ell(m(X^{(j)}), Y)}-\e{\ell(m(X), Y)}$ & Yes  \\
\hline
\end{tabular}

    \label{tab:th_ind}
\end{table*}

Next, in table \ref{tab:estim} we summarize the main methods and how they are estimated. We use the abbreviations: \textbf{P} for perturbation, \textbf{M} for marginalization, and \textbf{R} for refitting. To maintain readability in the table, we report the theoretical quantity rather than the explicit estimation formulas. In practice, these correspond to plug-in estimators: expectations are approximated using empirical means over a test set, and the function $m$ is replaced by a trained machine learning model.

For the construction of $X^{(j)}$, the $j$-th column is permuted across the dataset. The conditional version, $\widetilde{X}^{(j)}$, is obtained via a conditional sampler. In CPI-based methods, this conditional permutation is used, although any conditional sampler (as in CFI) is theoretically valid.

To distinguish between marginalization and refitting when estimating the restricted model $m_S$, we denote marginalization-based approaches as $\mathbb{E}[m(X) \mid X^S]$ (conditional sampling, then we use $v$), and refitting-based approaches simply as $m_S$. For marginal sampling, we denote $\mathbb{E}[m(X^{(-S)}) \mid X^S]$ (then, we use $v^m$).

\begin{table*}
 \centering
 \caption{\textbf{Summary of the VIMs:} The theoretical quantity targeted by the variable importance measures, the corresponding index, and the estimation procedure considered (M for marginalization, R for refitting, and P for perturbation). }
 \begin{tabular}{cccc}
\hline
\textbf{Method} & \textbf{Theoretical quantity} & \textbf{Index} & \textbf{Estimation} \\
\hline
cSAGE & $\sum_{S\subset -\{j\}}w_S\left(v(S\cup \{j\})-v(S)\right)$ & $\psi_\mathrm{SAGE}$ & M \\
\hline
cSAGEvf & $v(\{j\})$ & $\psi_\mathrm{LOCI}$ & M \\
\hline
mSAGEvf & $v^m(\{j\})$ & $\psi_\mathrm{mSAGEvf}$ & M \\
\hline
mSAGE & $\sum_{S\subset -\{j\}}w_S\left(v^m(S\cup\{j\})-v^m(S)\right)$ & $\psi_\mathrm{mSAGE}$ & M  \\
\hline
scSAGE & $v(-\{j\}\cup \{j\})-v(-\{j\})$ & $\psi_\mathrm{TSI}$ & M \\
\hline
LOCO & $\e{\ell(m_{-j}(X^{-j}), Y)}-\e{\ell(m(X), Y)}$ & $\psi_\mathrm{TSI}$ & R \\
\hline
LOCO-W & $\e{\ell(m_{-j}(X^{-j}), Y)}-\e{\ell(m(X), Y)}$ & $\psi_\mathrm{TSI}$ & R \\
\hline
LOCI & $\e{\ell(m_{j}(X^{j}), Y)}-\e{\ell(m(X), Y)}$ & $\psi_\mathrm{LOCI}$ & R \\
\hline
PFI & $\e{\ell(m(X^{(j)}, X^{-j}), Y)}-\e{\ell(m(X), Y)}$ & $\psi_\mathrm{PFI}$ & P \\
\hline
CFI & $\e{\ell(m(\widetilde{X}^{(j)}, X^{-j}), Y)}-\e{\ell(m(X), Y)}$ & $\psi_\mathrm{TSI}$ & P \\
\hline
Sobol-CPI(n-cal) & $\frac{n_\mathrm{cal}}{n_\mathrm{cal}+1}\left(\e{\ell(\frac{1}{n_\mathrm{cal}}\sum_{k=1}^{n_\mathrm{cal}}m(\widetilde{X}_k^{(j)}, X^{-j}),Y)}-\e{\ell(m(X), Y)}\right)$ & $\psi_\mathrm{TSI}$ & P/M \\
\hline
\end{tabular}
    \label{tab:estim}
\end{table*}

The difference between LOCO \citep{lei2018} and LOCO-W \citep{williamson2023general} lies in the use of a extra data splitting in the latter. This data splitting avoids using the same training set for both model refittings, which ensures that the variance does not vanish and enables asymptotic normality, thereby allowing for valid type-I error control. However, \cite{lobo2025doublerobustnessconditionalfeature} discuss the inference consequences of such data splitting. Principally, this causes the variance to explode, which leads to far fewer discoveries. This is also seen in our experiments (Section \ref{sec:exp}).

Sobol-CPI(n-cal) can be interpreted as a perturbation approach. Indeed, for $\text{n-cal} = 1$, it reduces to a normalized CPI/CFI. However, for larger values of n-cal, then marginalizing, it accounts for the estimation bias introduced by not having access to the true conditional expectation, which is not addressed by other marginalization-based methods.

In general, refitting-based approaches incur higher computational costs, as the models used to represent the relationship with the output tend to be more complex than those used for conditional sampling. Additionally, Shapley-based approaches are often slower, as they involve an exponential number of feature combinations. These factors should be taken into account when choosing an estimation procedure.


\section{Experiments}\label{sec:exp}

In this section, we compare all the presented global variable importance measures on both simulated and real datasets to illustrate two main points: (1) whether the theoretical index satisfies the minimal axiom—and therefore whether the estimate tends to zero when a covariate has no predictive power—and (2) the necessity of the general pipeline. Indeed, the estimated importance (and in particular the resulting ranking, which reflects the predictive power of each feature) remains consistent across estimation approaches, whether based on refitting, perturbation, or marginalization.

\subsection{Methods}
We compare \texttt{Sobol-CPI(1)} (a normalization of the \texttt{CPI} from \cite{chamma2023statistically}) and \texttt{Sobol-CPI(100)} from \cite{lobo2025doublerobustnessconditionalfeature}, \texttt{PFI} from \cite{Breiman2001}, \texttt{CFI} from \cite{Strobl2008}, marginal / conditional SAGE (\texttt{m/c SAGE}) and their respective value function variants (\texttt{m/c SAGEvf}) from \cite{covert2020understanding}, \texttt{scSAGEj} from \cite{Ewald_2024}, \texttt{LOCO} from \cite{lei2018}, \texttt{LOCO-W} from \cite{williamson2023general}, and \texttt{LOCI} (Leave One Covariate In). To enable comparison across all methods, the importance scores have been normalized. 
Most methods were computed using the Python package \textbf{fippy} (\url{https://github.com/gcskoenig/fippy}).
Code to reproduce these experiments is available at \url{https://github.com/AngelReyero/Principled-VIM-comparison}. 

The model used throughout the main experiments is a Gradient Boosting. However, in Section~\ref{sect:add_exp}, we repeat the same experiments using a Random Forest, Support Vector Regression (SVR), and a Neural Network, showing that the results are qualitatively similar.

All experiments were repeated at least 50 times using different random seeds. It is important to note that the boxplots illustrate the variability induced by the model optimization process across the seeds; they should not be interpreted as confidence intervals. Moreover, as discussed above, for many methods, such as PFI, neither statistical tests nor confidence intervals are available. Therefore, although a method may satisfy the minimal axiom, this remains a theoretical property of the corresponding importance index and does not, by itself, provide statistical guarantees.

\subsection{Simulated data}


We generate the input $X$ as a Gaussian vector with zero mean and a Toeplitz covariance matrix defined by $\Sigma_{i,j} = 0.6^{|i-j|}$, with dimension $p = 10$. The output is defined as $Y = X_0 + 2X_1 - X_4^2 + X_7 X_8$. We use $n = 5000$ samples for the main experiments. 
  
\begin{figure*}[htbp]
        \includegraphics[width=0.9\textwidth]{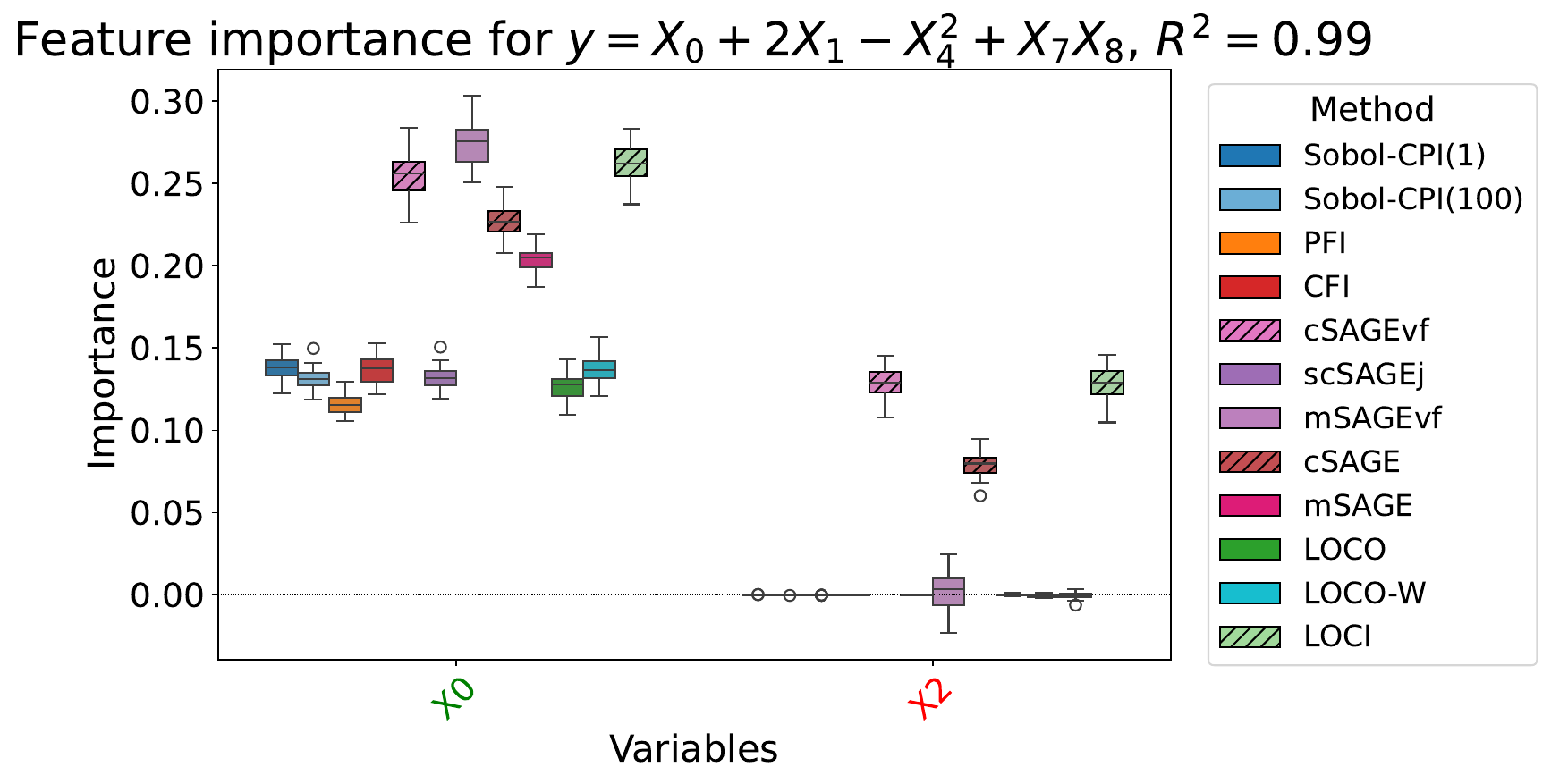}

    \caption{\textbf{Boxplot of estimated VIMs for an important and an unimportant variable:} Methods not satisfying theoretically the minimal axiom have boxes filled with diagonal hatch lines. The left panel shows an important covariate, while the right panel shows an unimportant one. Only conditional SAGE(vf) and LOCI, which
fail to satisfy the minimal axiom, assign non-zero importance to the unimportant
variable. \texttt{Sobol-CPI(1)}, \texttt{Sobol-CPI(100)}, \texttt{CFI}, \texttt{scSAGEj}, \texttt{LOCO}, and \texttt{LOCO-W} aim to estimate $\psi_\mathrm{TSI}$.
}
    \label{fig:reduced_boxplot}
\end{figure*}

In Figure~\ref{fig:reduced_boxplot}, we plot the estimated importances across all methods. For clarity, we present only one important and one unimportant feature, while the full set of estimated importances—from which similar conclusions can be drawn—is provided in Appendix \ref{app:sim_data}.

On the left of Figure~\ref{fig:reduced_boxplot}, for an important covariate, we observe that even though the quantities have been estimated using different inference methods, those corresponding to the same theoretical index yield similar importance values. On the right, there is the importance estimated to a not important coordinate. We observe that all the VIMs that satisfy the minimal axiom, represented by plain boxplots, provide a null importance. For instance, this illustrates that PFI is suited for conditional testing, contrary to what was said before. 

In Section \ref{sect:exp_convergence}, we examine how the estimation varies as $n$ ranges from 100 to 5000. The conclusions remain similar: numerical comparisons should be made between estimates of the same index to assess different inference properties, while conceptual comparisons should rely on their theoretical counterparts.

\subsection{Real data}
Following \cite{Ewald_2024}, we use the bike dataset from \cite{Fanaee-T2014}. 
In the main text, we present the results for Gradient Boosting. To show that the same conclusions hold more generally, Appendix~\ref{app:bike} presents the corresponding results for Random Forest, SVR, and Neural Networks.

\begin{figure*}[htbp]
        \includegraphics[width=0.9\textwidth]{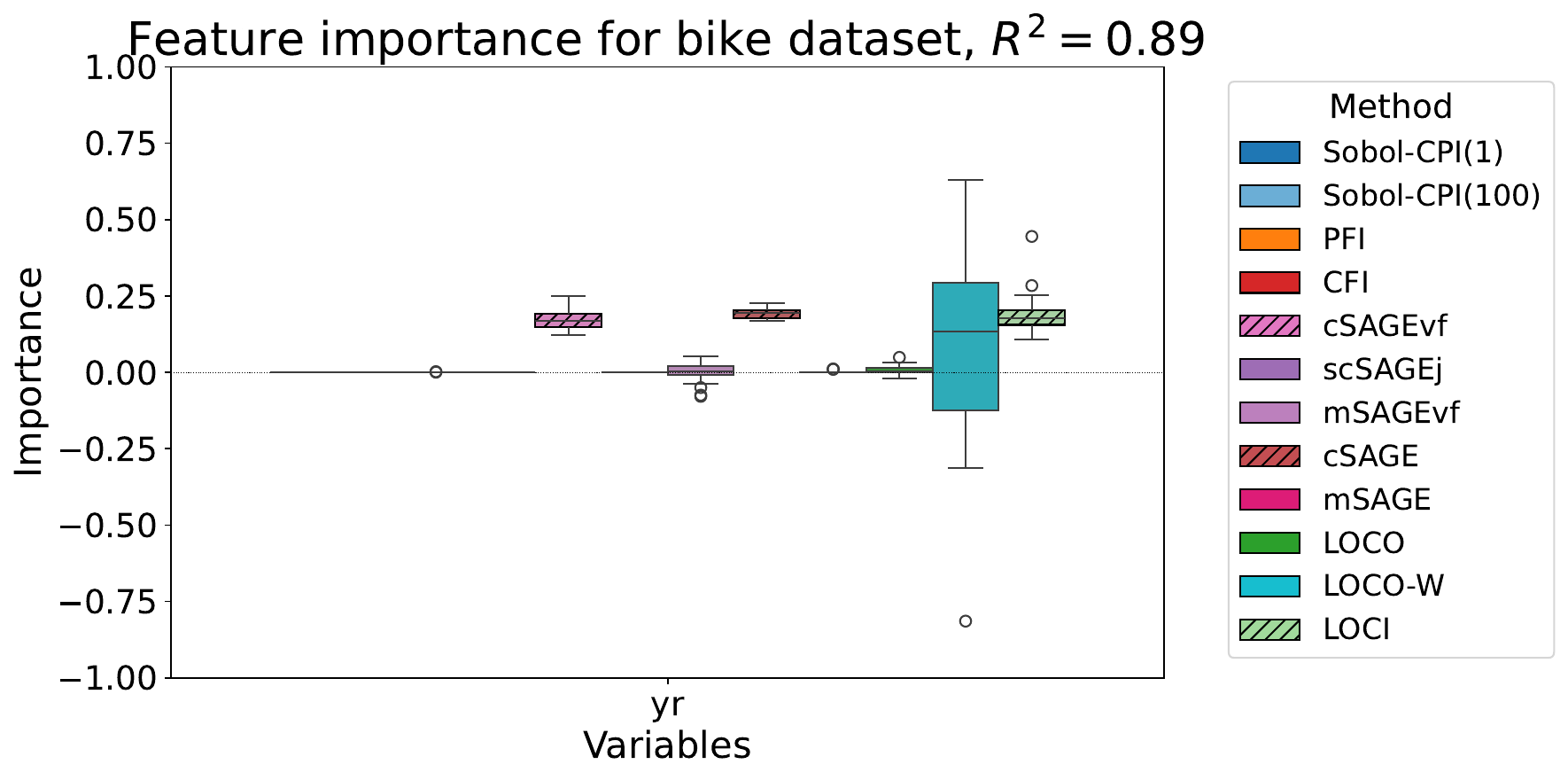}

    \caption{\textbf{Boxplots of the VIMs for the \textit{year} feature:} Methods not satisfying theoretically the minimal axiom have boxes filled with diagonal hatch lines. All methods that estimate an importance index satisfying the minimal axiom assign no importance to this variable.
}
    \label{fig:bike_main_1feature}
\end{figure*}

In Figure \ref{fig:bike_main_1feature}, we plot the estimated importance of the variable \textit{year} across all the discussed VIMs. We argue that this covariate is not important, i.e., it does not contribute predictive power to the model. Nevertheless, we observe that VIMs that do not satisfy the minimal axiom assign importance to this irrelevant covariate. Specifically, $\psi_\mathrm{SAGE}$ (estimated by cSAGE) and $\psi_\mathrm{LOCI}$ (estimated by cSAGEvf via marginalization and LOCI via refitting) both assign a nonzero importance. 
%


The claim of \cite{Ewald_2024} that PFI can assign high FI values to features even if they are not associated with the target but with other features that are associated with the target does not appear to be supported by the results of this experiment.
In fact, we observe in Figure \ref{fig:bike_main_1feature} that PFI, which satisfies the minimal axiom, assigns zero importance to the variable \textit{year}. This indicates that the model acts as a filter for conditional independence. As a result, PFI assigns different importance scores (and therefore different rankings) to the relevant covariates compared to methods like LOCO; however, both approaches identify the same set of important covariates.
Moreover, one may argue that the variable \texttt{year} is genuinely important but is simply not identified as such by any of the methods across the different models. To address this concern, we performed a similar experiment by introducing an artificial null feature that is correlated with the original predictors. The results lead to the same conclusions, as shown in Figure~\ref{fig:spurious_bike}.

\begin{figure*}[htbp]
        \includegraphics[width=0.9\textwidth]{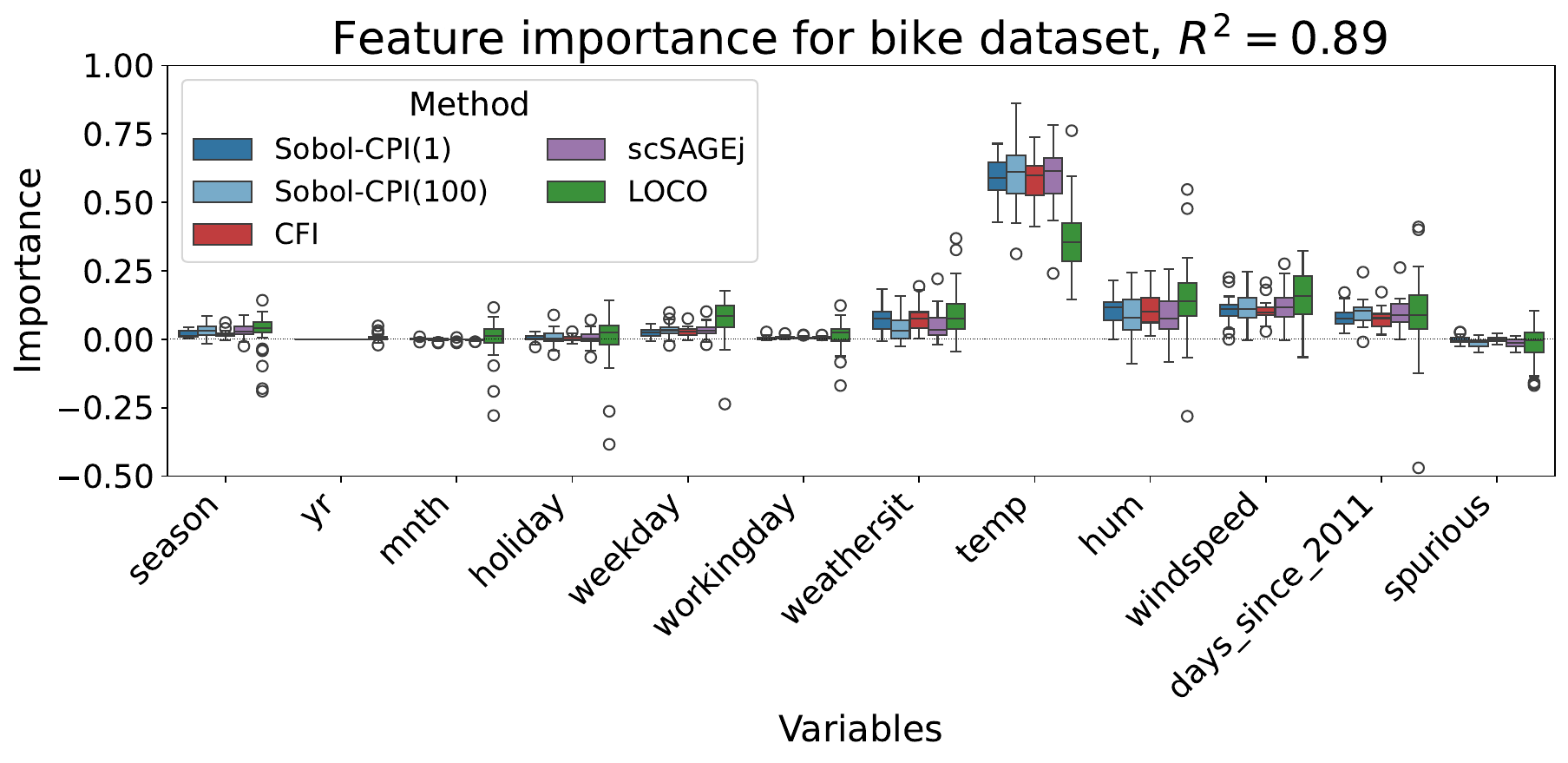}
    \caption{\textbf{Boxplots of the VIMs estimating $\psi_\mathrm{TSI}$ for all features:} All methods aim to estimate the same theoretical quantity. While their estimates are generally close, the refitting-based approaches exhibit poorer inference properties.
}
    \label{fig:bike_main_TSI}
\end{figure*}

In Figure \ref{fig:bike_main_TSI}, we present boxplots of all features using only the VIMs that aim to estimate the total Sobol' index. We omit \texttt{LOCO-W} due to its high variability, which compromises readability. 
We observe that all displayed methods produce similar importance values and rankings, even on real data, regardless of whether they rely on marginalization, refitting, or perturbation. This similarity arises because they target the same underlying quantity. However, their estimation properties differ. For instance, refitting-based methods exhibit greater variability \citep{paillard2025measuringvariableimportanceheterogeneous}, which can lead to inaccurate variable importance estimates that are highly sensitive to the data and the optimization procedure. Consequently, the resulting rankings tend to be less reliable in practice compared to those obtained via marginalization or perturbation. Furthermore, perturbation methods possess a double robustness property \citep{lobo2025doublerobustnessconditionalfeature}, which in practice translates into greater robustness to model misspecification under the null hypothesis, leading to faster convergence rates and reduced variance, as observed for variables such as \emph{month} and \emph{year}.

\section{Discussion}

In this article, we have proposed a new categorization of variable importance measures (VIMs). In doing so, we aim to provide practitioners with more insight into how to proceed when working with real datasets. We argue that comparing VIMs only based on their inference procedures it is not meaningful, and can be misleading. Instead, it is crucial to clearly define the theoretical index corresponding to the intended goal, and then select an appropriate estimation method. In particular, we emphasize that comparing methods aimed at estimating different indices is not meaningful, as each method yields a different ranking aligned with its own objectives.

This principled framework acknowledges that global variable importance aims to capture information about the true underlying distribution, with the model acting as a surrogate that simplifies this information. Furthermore, establishing a connection between variable importance and variable selection is essential for making insightful discoveries with statistical guarantees. Since the index is a theoretical quantity that must be estimated from data, it is crucial to provide statistical guarantees for the selected covariates. This connection is formalized through the minimal axiom, highlighting that both fields pursue the same objective.

The proposed axiom not only formalizes the notion of importance as predictive power but also generalizes it by relating it to the conditional independence testing framework under the established assumptions. We note that some notions of importance do not fall within this definition; for example, a feature may influence the output only through its variance without providing any predictive signal. However, we argue that when performing scientific inference using an ML model to explain the $X$--$Y$ relationship, no stronger notion can be achieved, since such alternative forms of importance are not captured by the predictive model.

Although the presented work provides a rigorous framework for developing VIMs, further research on this subject is necessary. First, new theoretical indices must be developed to capture alternative notions of importance with an explicit articulation of the axioms they are designed to satisfy. Additionally, more work is needed on the estimation side, particularly for challenging indices such as the decorrelated LOCO, for which accurate estimation is complicated by inference bias \cite{LocoVSShapley}. Moreover, high-dimensional data pose additional challenges, both in terms of inference, for which convergence and statistical guarantees may fail to hold \cite{Berrett, blain2025knockoffsfaildiagnosingfixing}, and computational feasibility \cite{covert2020understanding,pmlr-v130-covert21a}. These limitations can substantially restrict the applicability of such methods in complex, high-dimensional settings.




\begin{appendix}
\section{Notation Glossary}\label{app:glossary}
The notation used in this paper is gathered in Table~\ref{tab:notation}.
\begin{table}[htbp]
\centering
\caption{Notation used in the paper.}
\small  
\begin{tabular}{p{4cm} p{4cm}}
\textbf{} & \textbf{Description} \\
\hline
$X\in \mathbb{R}^p$ & Input \\
$j\in [p]$ & Feature of interest \\
$X^j\in \mathbb{R}$ & $j$-th input covariate \\
$X^{-j}\in \mathbb{R}^{p-1}$ & $X$ with the $j$-th covariate excluded \\
$X^S\in \mathbb{R}^{p-|S|}$ & $S$-th input covariates for $S\subset[p]$ \\
$Y\in \mathbb{R}$ & Output \\
$P_{(X, Y)}\in \mathcal{M}$ & Distribution of $(X, Y)$ \\
$\check{X}^{(j)}$ & General perturbation \\
$X^{(j)}$ & Marginal perturbation \\
$\widetilde{X}^{(j)}$ & Conditional perturbation \\
$m(X)$ (resp. $m_{-j}(X^{-j})$) & $\e{Y\mid X}$ (resp. $\e{Y\mid X^{-j}}$) \\
$m_S(X)$ & $\e{Y\mid X^S} = \e{m(X)\mid X^S}$ \\
$\widehat{m}$ (resp. $\widehat{m}_{-j}$) & Estimation of $m$ (resp. of $m_{-j}$) \\
$\ell$ & Loss function \\
$\mathcal{F}$ & Generic space of functions \\
$\psi(j, P)$ & Importance index of $j$ under $P$ \\
$n_\mathrm{cal}$ & Size of calibration set \\
\end{tabular}
\label{tab:notation}
\end{table}
%
 Also, we recall that marginal perturbation means 
 \[
P_{(X^{(j)}, X^{-j}, Y)} = P_{X^j} \times P_{(X^{-j}, Y)}.
\]
i.e. all the features are preserved but the $j$-th feature, which follows the same marginal as the original $P_{(X^{(j)})}=P_{(X^j)}$ and $X^{(j)}\indep X^{-j}, Y$. The conditional is $$P_{(\widetilde{X}^{(j)}, X^{-j}, Y)}=P_{X^j\mid X^{-j}}\times P_{(X^{-j}, Y)},$$ so all the features are preserved, and the $j$-th feature is conditionally independent of the output $\widetilde{X}^{(j)}\indep Y\mid X^{-j}$ and preserved the original distribution $P_{(\widetilde{X}^{(j)}, X^{-j})}=P_{(X)}$.

\section{Conditional and functional independence}

In this section, we prove that, under certain standard assumptions, it is possible to establish the equivalence between conditional independence and functional independence (Proposition \ref{prop:CIT_functional_dep}), i.e., between $Y\indep X^{j} \mid X^{-j}$ and $m(X) \in \mathcal{F}_{-j}$. 

First, under the standard additive noise assumption~(\ref{ass:additive_noise}), \cite{lobo2025doublerobustnessconditionalfeature} proved this equivalence in Lemma I.1 of the appendix, which is based on proving that $m(X)$ is $\sigma(X^{-j})$-measurable. 

For classification, we will establish the result under the general classification assumption~(\ref{ass:classif}).


First, if $Y\indep X^{j}\mid X^{-j}$, we have that 
\begin{align*}
    \sigma(m(X))&=\mathbb{P}(Y=1\mid X)\comment{using Assumption \ref{ass:classif}}
    \\
    &=\mathbb{P}(Y=1\mid X^{-j})\comment{using that $Y\indep X^{j}\mid X^{-j}$}
    \\&=f_{-j}(X^{-j}),
\end{align*} for some $f_{-j}\in \mathcal{F}_{-j}$, as $\mathbb{P}(Y=1\mid X^{-j})$ belongs to the $\sigma$-algebra generated by $X^{-j}$. 

Then, if $m\in \mathcal{F}_{-j}$, then we have that 
\begin{align*}
    \mathbb{P}(Y=1\mid X^{-j})&=\e{\mathbb{P}(Y=1\mid X)\mid X^{-j}}\comment{using the tower rule}\\
    &=\e{\sigma(m(X))\mid X^{-j}}\comment{using Assumption \ref{ass:classif}}\\
    &=\e{\sigma(m_{-j}(X^{-j}))\mid X^{-j}}\comment{using that $m\in \mathcal{F}_{-j}$}\\
    &=\sigma(m_{-j}(X^{-j}))\comment{$\sigma(m_{-j}(X^{-j}))$ is $\sigma(X^{-j})$-measurable}\\
    &=\sigma(m(X))\comment{using definition of $m_{-j}$}\\
    &=\mathbb{P}(Y=1\mid X).
    \end{align*}
    Therefore, as $Y$ binary and $\mathbb{P}(Y=1\mid X^{-j})=\mathbb{P}(Y=1\mid X)$, then $Y\indep X^{-j}\mid X^{-j}$.

\section{Equivalent reformulations of the Total Sobol’ Index}

For the sake of completeness, in this section we provide the explicit equalities derived for the Total Sobol' Index under the quadratic loss in Proposition~\ref{prop:tsi_reform}. Namely,
\firsteuclid*

Note in particular that there is no need to assume the additive noise assumption \ref{ass:additive_noise}; it suffices to use that $m(X):= \mathbb{E}[Y\mid X]$ and $m_{-j}(X^{-j}):= \mathbb{E}[Y\mid X^{-j}]$. Also, note that the equality in \eqref{eq:tsi_perturb} is a particular case of the Sobol-CPI from \cite{lobo2025doublerobustnessconditionalfeature} when $n_{\mathrm{cal}}=1$.

\begin{proof}
    
We start by proving that \eqref{eq:tsi_diff_model} equals \eqref{eq:tsi_squared}. 
\begin{align*}
    \mathbb{E}&\left[(Y-m_{-j}(X^{-j}))^2\right]-\e{(Y-m(X))^2}\\&= \e{(Y-m(X)+m(X)-m_{-j}(X^{-j}))^2}\\&\qquad-\e{(Y-m(X))^2}\\
    &= \e{(Y-m(X))^2}+\e{(m(X)-m_{-j}(X^{-j}))^2}  \\ &\qquad + 2\e{(Y-m(X))(m(X)-m_{-j}(X^{-j}))}\\&\qquad-\e{(Y-m(X))^2}\\
    &= \e{(m(X)-m_{-j}(X^{-j}))^2}\\& \qquad + 2\e{(Y-m(X))(m(X)-m_{-j}(X^{-j}))}.
\end{align*}
Now, note that using Tower's property we have that

\begin{align*}
    \mathbb{E}&\left[(Y-m(X))(m(X)-m_{-j}(X^{-j}))\right] \\&= \e{\e{(Y-m(X))(m(X)-m_{-j}(X^{-j}))\mid X}}\\&= \e{(\e{Y-m(X)\mid X})(m(X)-m_{-j}(X^{-j}))} = 0.
\end{align*}

To prove equivalence with \eqref{eq:tsi_marg}, note that 
\begin{align*}
    m_{-j}(X^{-j})&=\e{Y\mid X^{-j}}=\e{\e{Y\mid X}\mid X^{-j}}\\&=\e{m(X)\mid X^{-j}}.
\end{align*}

Finally, to prove \eqref{eq:tsi_perturb}, similarly as before, we have that

\begin{align*}
    \mathbb{E}&\left[(Y-m(\widetilde{X}^{(j)}, X^{-j}))^2\right]-\e{(Y-m(X))^2}
    \\&= \e{(Y-m(X)+m(X)-m(\widetilde{X}^{(j)}))^2}\\&\qquad-\e{(Y-m(X))^2}\\
    &= \e{(Y-m(X))^2}+\e{(m(X)-m(\widetilde{X}^{(j)}))^2}\\&\qquad+2\e{(Y-m(X))(m(X)-m(\widetilde{X}^{(j)}, X^{-j}))}\\&\qquad-\e{(Y-m(X))^2}\\
    &=\e{(m(X)-m(\widetilde{X}^{(j)}))^2}\\&\qquad+2\e{(Y-m(X))(m(X)-m(\widetilde{X}^{(j)}, X^{-j}))}.
\end{align*}

The last term cancels since

\begin{align*}
    &\mathbb{E}\left[(Y-m(X))(m(X)-m(\widetilde{X}^{(j)}, X^{-j}))\right]\\ &= \e{\e{(Y-m(X))(m(X)-m(\widetilde{X}^{(j)}, X^{-j}))\mid X, \widetilde{X}^{(j)}}}\\
    &= \e{(\e{Y-m(X)\mid X, \widetilde{X}^{(j)}})(m(X)-m(\widetilde{X}^{(j)}, X^{-j}))}\\
    &=\e{(\e{Y-m(X)\mid X})(m(X)-m(\widetilde{X}^{(j)}, X^{-j}))}=0.
\end{align*}

We conclude by noting that
\begin{align*}
    \mathbb{E}&\left[(m(X)-m(\widetilde{X}^{(j)}, X^{-j}))^2\right] \\&= \e{\e{(m(X)-m(\widetilde{X}^{(j)}, X^{-j}))^2\mid X^{-j}}}\\&=2\e{\mathrm{Var}(m(X)\mid X^{-j})},
\end{align*}
since $X^{j}\sim \widetilde{X}^{(j)}\mid X^{-j}$ and independent.

\end{proof}

\section{Minimal axiom proofs}\label{sect:ma_proofs}
In this section, we study whether the theoretical indices presented in the main text satisfy the minimal axiom.

\subsection{\texorpdfstring{Generalized Linear Models $\psi_\mathrm{GLM}$}{}}
\begin{proof}[Proof of Proposition \ref{prop:glm}]
On the one hand, assume that $X^j\indep Y\mid X^{-j}$. Then, we have that $g(X^1\beta_1+\ldots +X^p\beta_p)=:\e{Y\mid X}=\e{Y\mid X^{-j}}$, then it is $\sigma(X^{-j})$-measurable. However, as $X^j$ is not $\sigma(X^{-j})$-measurable by identifiability assumption, then $\beta_j=0$. 

On the other hand, assume that $\beta_j=0$. Then, in particular using the GLM likelihood assumption \citep{mccullagh1989generalized}:
$$p_Y(Y; \theta, \phi)=\mathrm{exp}\left(\frac{Y\theta-\mathrm{b}(\theta)}{\mathrm{a}(\phi)}+\mathrm{c}(Y;\phi)\right),$$
we have that the canonical parameter does not depend on $X^j$ because $\beta_j=0$, then $p_y(Y\mid X)=p_y(Y\mid X^{-j})$.
\end{proof}

\subsection{\texorpdfstring{Perturbation $\psi_\mathrm{perturb}$}{}}

\begin{proof}[Proof of Proposition \ref{prop:perturb_ma}]

First, we prove it in particular for the quadratic loss. For any perturbation $\widetilde{X}^j$, using Assumption \ref{ass:additive_noise} we have that 
\begin{align*}
    \psi_\mathrm{perturb}&(j,P)=\e{(Y- m(\widetilde{X}^j))^2}-\e{(Y- m(X))^2}\\&=
    \e{(Y- m(X^{-j}, \widetilde{X}^{j}))^2}-\e{(Y- m(X))^2}
    \\&=\e{(m(X)+\epsilon- m(X^{-j}, \widetilde{X}^{j}))^2}\\&\qquad-\e{(m(X)+\epsilon- m(X))^2}\\
    &=\e{(m(X)-m(X^{-j}, \widetilde{X}^{j}))^2}. 
\end{align*}
Thus, using the strict convexity of the quadratic loss, we have that $m(X) - m(X^{-j}, \widetilde{X}^{j}) = 0$ almost surely. 

More generally, for any strictly convex loss $\ell$, we have that $\e{\ell(Y, m(\widetilde{X}^j))}-\e{\ell(Y, m(X))}=0$ if and only if $m(X)=m(\widetilde{X}^j)$ almost surely. Note that, up to this point, we have not specified any particular perturbation. Therefore, we can conclude that $\psi_\mathrm{perturb}=0$ if and only if the Bayes predictor is invariant under the perturbation. Now, to establish functional independence and later conclude the minimal axiom using Proposition~\ref{prop:CIT_functional_dep}, we need to exploit specific properties of each perturbation.

The first perturbation considered is the constant perturbation $a\in\mathbb{R}$. In this case,
\[
m(X^j, X^{-j})=m(a, X^{-j}),
\]
which is clearly $\sigma(X^{-j})$-measurable and therefore functionally independent.

Next, consider the conditional perturbation. We have
\[
m(X)=m(\widetilde{X}^{(j)}, X^{-j}) \quad \text{a.s.},
\]
where $P_{(\widetilde{X}^{(j)}, X^{-j})}=P_X$ and $\widetilde{X}^{(j)}$ is conditionally independent given $X^{-j}$. Therefore, by integrating with respect to $\widetilde{X}^{(j)}$ conditional on $X^{-j}$, we obtain
\begin{align*}
    m(X)
    &=\mathbb{E}_{\widetilde{X}^{(j)}\mid X^{-j}, X^j}[m(X)]\\
    &=\mathbb{E}_{\widetilde{X}^{(j)}\mid X^{-j}, X^j}[m(\widetilde{X}^{(j)}, X^{-j})]\\
    &=\mathbb{E}_{\widetilde{X}^{(j)}\mid X^{-j}}\!\left[\mathbb{E}[Y\mid \widetilde{X}^{(j)}, X^{-j}]\right]\\
    &=\mathbb{E}[Y\mid X^{-j}].
\end{align*}

Finally, consider the marginal perturbation $X^{(j)}$ used in PFI. We need to show that $m(X^j,X^{-j})$ is almost surely constant in $X^j$, conditional on $X^{-j}$. Indeed, suppose this were not the case. Then there would exist a measurable set $A$ such that
\[
0<\mathbb{P}\bigl(m(X^j,X^{-j})\in A\mid X^{-j}\bigr)<1.
\]
Hence,
\[
\mathbb{P}\bigl(m(X^{(j)},X^{-j})\notin A\mid X^{-j}\bigr)>0.
\]
Since $X^{(j)}$ is conditionally independent of $X^j$ given $X^{-j}$,
\begin{align*}
\mathbb{P}&\bigl(m(X^j,X^{-j})\in A,\,
m(X^{(j)},X^{-j})\notin A
\mid X^{-j}\bigr)
\\&=
\mathbb{P}\bigl(m(X^j,X^{-j})\in A\mid X^{-j}\bigr)\\ &\quad\times
\mathbb{P}\bigl(m(X^{(j)},X^{-j})\notin A\mid X^{-j}\bigr)
>0,
\end{align*}

which contradicts the perturbation invariance
\[
m(X^j,X^{-j})=m(X^{(j)},X^{-j}) \quad \text{a.s.}
\]
Note that the same argument could also have been used for the conditional perturbation.


For the second part of the proposition, it suffices to apply Theorem 3.1 from \cite{lobo2025doublerobustnessconditionalfeature}. Indeed, this result follows from one of the double robustness properties of CFI: for any perturbation, using the asymptotic relevance of $\widehat{m}$, the importance vanishes.
 
\end{proof}

\subsubsection{PFI is not suitable for marginal testing.}\label{subsect:PFI_marginal}

The marginal construction of the permutation in PFI \citep[see][]{Ewald_2024} might suggest that it is only suited for marginal testing. In this section, we show that this is not the case, in the sense that the theoretical index estimated by PFI in fact aligns with conditional, not marginal, independence. First, using Proposition~\ref{prop:perturb_ma}, we show that if PFI is nonzero, then $X^j$ is dependent on the output given the rest of the input. Then, we demonstrate that the assumptions made in \cite{Ewald_2024} to draw conclusions about marginal independence also imply conditional independence. Finally, we present counterexamples illustrating the limitations of PFI as a measure of marginal independence, showing cases in which the PFI is zero despite marginal dependence, and conversely, cases in which it is nonzero despite marginal independence.

\paragraph{PFI different from 0:}
\citet[Result 5.11]{Ewald_2024} proved that if $\psi_\mathrm{PFI}(j, P)\neq 0$ and $X^j\indep X^{-j}\mid Y$, then  $X^j\notindep Y$. In fact, Proposition \ref{prop:perturb_ma} implies that if $\psi_\mathrm{PFI}(j, P)\neq 0$ then we have that $X^j\notindep Y\mid X^{-j}$. (More generally, this applies to any perturbation index). The problem arises from the fact that the model used in the index is not an arbitrary model, but rather an estimate of the data-generating process. Under standard assumptions, this model acts as a filter for conditional independence. Consequently, if the difference in loss is nonzero, it implies that the model relied on the covariate, and therefore, the covariate is important—i.e., it is conditionally dependent.

\paragraph{PFI equal to 0:}
\citet[Result 5.11]{Ewald_2024}  proved that for the cross-entropy loss if $\psi_\mathrm{PFI}(j, P)=0, X^j\indep X^{-j}$ and $X^j\indep X^{-j}\mid Y$, then $X_j\indep Y$. However, this also implies that $X^j\indep Y\mid X^{-j}$. Indeed, we have that
\begin{align*}
    p(X^j, Y\mid X^{-j})&=\frac{p(X^j, Y, X^{-j})}{p(X^{-j})}
    \\&= \frac{p(Y)p(X^{j}\mid Y)p(X^{-j}\mid Y)}{p(X^{-j})}
    \\&= \frac{p(Y)p(X^{j})p(X^{-j}\mid Y)}{p(X^{-j})}\\
    &=\frac{p(X^{j})p(X^{-j}, Y)}{p(X^{-j})}\\&= \frac{p(X^{j})p(Y\mid X^{-j})p(X^{-j})}{p(X^{-j})}\\&= p(X^{j})p(Y\mid X^{-j}) \\
    &= p(X^{j}\mid X^{-j})p(Y\mid X^{-j}).
\end{align*}
Therefore, we conclude that $X^j\indep Y\mid X^{-j}$.

In general, as shown in Proposition~\ref{prop:perturb_ma}, there is no need to impose such restrictive assumptions to draw conclusions about conditional independence. We present this development because \cite{Ewald_2024} first showed that if a method is suited for detecting marginal independence, then it is not suited for detecting conditional independence. Second, they showed that under this restrictive setting, PFI is appropriate for marginal independence. We therefore demonstrate that, in this particular case, it also implies conditional independence, thus aligning with our more general result.

\paragraph{PFI marginal counterexamples.}
In this section, we present counterexamples to the interpretation of PFI as a measure of marginal importance. In particular, we provide examples in which the features are marginally independent while the PFI is nonzero, as well as examples in which the PFI is zero despite marginal dependence.

The latter is easily illustrated by the setting in Figure~\ref{fig:margin_pfi_count1}. Suppose that $X^j$ is correlated with $X^{-j}$ and that $Y$ depends only on $X^{-j}$. Then, conditional independence implies that the Bayes predictor does not depend on $X^j$,  and, consequently, a PFI equal to zero. However, since $X^j$ is correlated with $X^{-j}$, it is still marginally dependent on $Y$.

\begin{figure}
\centering
\begin{tikzpicture}[
    node distance=0.75cm,
    every node/.style={draw, circle, minimum size=1cm}
]
\node (X) {$X^j$};
\node (Z) [right=of X] {$X^{-j}$};
\node (Y) [above=of Z] {$Y$};

\draw[->] (X) -- (Z);
\draw[->] (Z) -- (Y);

\end{tikzpicture}
\caption{$X^j\notindep Y$ but $\psi_\mathrm{PFI}(j)=0$.}\label{fig:margin_pfi_count1}
\end{figure}

Figure~\ref{fig:margin_pfi_count2} illustrates the opposite situation, where the features are marginally independent but the PFI is nonzero because they are not conditionally independent and therefore still provide predictive information. For example, let $X^j$ and $Y$ be sampled independently from $\mathrm{Bern}(0.5)$, and define
\[
X^{-j}=Y\oplus X^j,
\]
where $\oplus$ denotes the exclusive OR operation. Then, $X^j$ and $Y$ are marginally independent but not conditionally independent. Consequently, the Bayes predictor depends on the $j$-th feature, and therefore $\psi_{\mathrm{PFI}}\neq 0$.


\begin{figure}
\centering
\begin{tikzpicture}[
    node distance=0.75cm,
    every node/.style={draw, circle, minimum size=1cm}
]
\usetikzlibrary {arrows.meta} 
\node (X) {$X^j$};
\node (Z) [right=of X] {$X^{-j}$};
\node (Y) [above=of Z] {$Y$};

\draw[->] (X) -- (Z);
\draw[->] (Y) -- (Z);

\end{tikzpicture}
\caption{$X^j\indep Y$ but $\psi_\mathrm{PFI}(j)>0$.}\label{fig:margin_pfi_count2}
\end{figure}

\subsection{\texorpdfstring{General Total Sobol' Index $\psi_\mathrm{TSI}$}{}}


\begin{proof}[Proof of Proposition \ref{prop:TSI_ma}]
    \cite{covert2020understanding} established this result in the context of the cross-entropy loss. Specifically, they showed that
\[
\mathrm{I}(Y; X^j \mid X^{-j}) = 0 \quad \Longleftrightarrow \quad X^j \perp\!\!\!\perp Y \mid X^{-j},
\]
which corresponds precisely to the minimal axiom using \eqref{eq:TSI_CE}.

For the quadratic loss, we can directly use the CFI formulation of the TSI and apply Proposition \ref{prop:perturb_ma}.

Under the assumptions of the proposition, this equivalence follows directly. Since the loss $\ell$ is strictly convex, it admits a unique minimizer, which by definition is the Bayes predictor, a function of $m\in \mathcal{F}$. Hence, we have that
\begin{align*}
\mathbb{E}[\ell(Y, m(X))] = \mathbb{E}[\ell(Y, m_{-j}(X^{-j}))] \\\quad \Longleftrightarrow \quad m(X) = m_{-j}(X^{-j}) \text{ a.s.}
\end{align*}
This condition implies that $X^j$ provides no additional predictive power beyond $X^{-j}$, i.e., $m$ is functionally independent from $X^j$ given $X^{-j}$. By the equivalence between functional and conditional independence under these assumptions, we conclude that $\psi_{\mathrm{TSI}}(j, P)$ satisfies the minimal axiom. 

\end{proof}

\subsection{\texorpdfstring{Shapley Additive Global importancE (SAGE, $\psi_\mathrm{SAGE}$)}{}}

\begin{proof}[Proof of Proposition \ref{prop:SAGE}] We begin by showing that $\psi_\mathrm{SAGEvf}$ does not satisfy the minimal axiom, as there exists a spurious correlation between the $j$-th coordinate and an important coordinate. Similarly, since $\psi_\mathrm{SAGE}$ combines multiple terms—including those reflecting this spurious importance—it also fails to satisfy the minimal axiom. Finally, we prove that $\psi_\mathrm{scSAGEj}$ coincides with $\psi_\mathrm{TSI}$ and, in particular, as established in the previous section, satisfies the minimal axiom.
    
We recall the notation $m_S(X^{S}) = \mathbb{E}[Y \mid X^{S}] = \mathbb{E}[m(X) \mid X^{S}]$ for $S \subseteq [p]$. In particular, for $S = \{j\}$, we denote it simply as $m_j$.
    
    We start with the SAGE value function. Indeed, it is given by 
    \begin{align*}
        \psi_{\mathrm{SAGEvf}}(j, P):=\e{\ell(\e{Y}, Y)}-\e{\ell(m_j(x^j), Y)},
    \end{align*}
    with $m_j(x^j):=\e{m(X^j, X^{-j})\mid X^j=x^j}$ $=\mathbb{E}[Y\mid X^j=x^j]$. Therefore, it is easy to construct the following general counterexample: for $Y=m_{-j}(X^{-j})+\epsilon$, with $\epsilon\indep X$, we observe that $Y\indep X^{j}\mid X^{-j}$, but if $X^{j}\cancel{\indep}X^{-j}$, it will not be $0$ as $\e{Y\mid X^j}\neq \e{Y}$.
    
    For the SAGE, it is given by
    \begin{align*}
        \psi_\mathrm{SAGE}(j, P)=\frac{1}{d}\sum_{S\subset -\{j\}}\binom{d-1}{|S|}^{-1}\left(v(S\cup \{j\})-v(S)\right),
    \end{align*}
    with $v(S)=\e{\ell(Y, \e{Y})}-\e{\ell(Y, m_S(x^S)}$. In particular, using the same counterexample as for the SAGE value function, we observe that for $S = \emptyset$ the difference is strictly positive as $v(\emptyset\cup \{j\})-v(\emptyset)=v(\{j\})>0$. Therefore, $\psi_\mathrm{cSAGE}(j, P) > 0$, since it is a combination of differences that are either positive or zero, with at least one strictly positive. However, since $X^j \indep y \mid X^{-j}$, the minimal axiom is not satisfied.

    For the $\psi_\mathrm{scSAGEj}$, we note that it is exactly the total Sobol' Index:
    \begin{align*}
        \psi_\mathrm{scSAGEj}&:=v(-\{j\}\cup \{j\})-v(-\{j\})\\
        &= \e{\ell(Y, \e{Y})}-\e{\ell(Y, m_{-\{j\}\cup\{j\}}(x^{-\{j\}\cup \{j\}})}\\&\qquad-\e{\ell(Y, \e{Y})}+\e{\ell(Y, m_{-\{j\}}(x^{-\{j\}})}\\
        &= \e{\ell(Y, m_{-\{j\}}(x^{-\{j\}})}-\e{\ell(Y, m(x)}\\&=\psi_{\mathrm{TSI}}(j, P). 
    \end{align*}
    In particular, it satisfies the minimal axiom (see Proposition \ref{prop:TSI_ma}). 
\end{proof}

\subsection{\texorpdfstring{Marginal SAGE $\psi_\mathrm{mSAGE}$}{}}

\begin{proof}[Proof of Proposition \ref{prop:mSAGE}]

    For the notation, we denote 
\[
m^m_S(X^S) := \mathbb{E}[m(X^{(-S)}) \mid X^{S}].
\]
This means, in particular, that the coordinates in $X^{(-S)S} := X^S$ are fixed, and the expectation is taken over the remaining coordinates, which are independent from $X^{S}$ due to marginal permutation.

We observe that, for this proof, there is no need to preserve any specific distribution over the variables on which the expectation is taken (i.e., $X^{(-S)-S}$); it suffices to have independence. In this way, the coordinates over which we take the expectation are not influenced by the others, and thus the conditional filtering of the trained function remains valid—unlike in the original SAGE.

To make this point more explicit, note that by independence,
\begin{align*}
m^m_S(x^S) &:= \mathbb{E}[m(X^{(-S)}) \mid X^{S} = x^S] \\&= \mathbb{E}[m(X^{(-S)-S}, x^S) \mid X^{S} = x^S] \\&= \mathbb{E}[m(X^{(-S)-S}, x^S)].
\end{align*}
Therefore, the information provided by the coordinates in $S$ does not influence the distribution of the remaining coordinates.

    If the sampling is done marginally, we have in particular that $m^m_S(x^S)=\mathbb{E}[m(X^{-S}, x^S)]$.
    
    We start with the marginal SAGE value function. It is defined as 
    \begin{align*}
        \psi_{\mathrm{mSAGEvf}}(j, P):=\e{\ell(\e{Y}, Y)}-\e{\ell(m_j^m(x^j), Y)}.
    \end{align*}
    Note that $m_j^m(x^j):=\e{m(x^j, X^{-j})}$. Under the null hypothesis, $m\in \mathcal{F}_{-j}$, and therefore $$m_j^m(x^j)=\mathbb{E}[m(x^j, X^{-j})] =\e{m_{-j}(X^{-j})}=\e{Y}.$$ Thus, its value is 0 if and only if $x^j$ is conditional independent. Similarly, for the marginal SAGE, it can be written as 
    \begin{align*}
        \psi_\mathrm{mSAGE}=\frac{1}{d}\sum_{S\subset -\{j\}}\binom{d-1}{|S|}^{-1}\left(v^m(S\cup \{j\})-v^m(S)\right),
    \end{align*}
    with $v^m(S)=\e{\ell(Y, \e{Y})}-\e{\ell(Y, m^m_S(x^S)}$. Thus, we just need to note that $v^m(S\cup \{j\})=v^m(S)$ for all $S\subset -\{j\}$. To see this, similarly as before, using the equivalence between conditional dependence and functional dependence, we observe that 
    \begin{align*}
        m^m_{S\cup \{j\}}(x_{S\cup\{j\}})&=\e{m(x^{S\cup\{j\} }, X^{-(S\cup \{j\})})}\\&=\e{m_{-j}(x^{S}, X^{-(S\cup \{j\})})}\\&=\e{m(x^{S }, X^{-S\cup\{j\}})}\\&= m^m_{S}(x_{S}).
    \end{align*}
    Then, we have that 
    \begin{align*}
        v^m(S\cup\{j\})&=\e{\ell(Y, \e{Y})}-\e{\ell(Y, m^m_{S\cup\{j\}}(x^{S\cup \{j\}})}\\&=\e{\ell(Y, \e{Y})}-\e{\ell(Y, m^m_S(x^S)}\\&=v^m(S).
    \end{align*}
\end{proof}

\subsection{\texorpdfstring{Leave One Covariate In (LOCI, $\psi_\mathrm{LOCI}$)}{}}\label{sect:LOCI}

\begin{propo}\label{prop:LOCI}
$\psi_\mathrm{LOCI}$ does not satisfy the minimal axiom.
\end{propo}

\begin{proof}[Proof of Proposition \ref{prop:LOCI}]
    Similarly to the (conditional) SAGE value function for the subset $\{j\}$, we have that it is given by
\begin{align*}
    \mathbb{E}[\ell(Y, \mathbb{E}[Y])] - \mathbb{E}[\ell(Y, m_j(X^j))],
\end{align*}
except that the index $j$ appears as a refitting index for the estimation rather than a marginalization index. Nevertheless, the index remains the same, and the same counterexample applies to show that this difference is non-zero, even under conditional independence.

\end{proof}

\section{Asymptotic relevance assumption}\label{sect:asymp_rel}

Under the conditional null hypothesis, i.e.\ $X^j \indep Y \mid X^{-j}$, the Bayes predictor does not depend on $X^j$, since this feature is not predictive. However, since the predictor must be estimated from data, it is natural to ask whether the estimated model depends asymptotically on this coordinate. This is formalized through the asymptotic relevance assumption \ref{ass:asymp_relevance}, introduced by \cite{lobo2025doublerobustnessconditionalfeature}. They discuss this assumption for several model classes, and prove that if the loss function is strictly convex (as in the case of the squared loss), then for any model $f$,

\begin{align*} \mathbb{E}\!\left[\ell\bigl(f(X^{-j}, X^j), Y\bigr)\right] &= \mathbb{E}\!\left[ \mathbb{E}\!\left[ \ell\bigl(f(X^{-j}, X^j), Y\bigr) \,\middle|\, X^{-j}, Y \right] \right] \\ &\geq \mathbb{E}\!\left[ \ell\!\left( \mathbb{E}\!\left[f(X^{-j}, X^j)\mid X^{-j}, Y\right], Y \right) \right] \\ &= \mathbb{E}\!\left[ \ell\!\left( \mathbb{E}\!\left[f(X^{-j}, X^j)\mid X^{-j}\right], Y \right) \right], \end{align*}

where the last equality follows from the conditional independence. Equality holds if and only if $f(X^{-j}, X^j)$ is $\sigma(X^{-j})$-measurable, which implies that $f$ is almost surely constant in $X^j$ and therefore does not depend on this feature. Consequently, a model that ignores $X^j$ can achieve a strictly smaller risk. Note that standard consistency results for empirical risk minimization prove that the estimated model's risk converges to the Bayes risk, which does not depend on $X^j$. Hence, it is reasonable to expect the learned model not to rely on this feature asymptotically.

In general, the convergence of the permutation feature importance (PFI) is not clear. Indeed, due to the permutation, the model is evaluated in regions of the space where it was not trained, leading to extrapolation bias \citep{Hooker2021}, and therefore the consistency of the plug-in estimator is not guaranteed. \cite{scornet2022mda} proved that, under some assumptions on the support, PFI converges for Random Forests. Note, however, that we are only claiming convergence under the null from a model that is already trained on the data. Therefore, if the model converges to the Bayes predictor, which itself does not depend on the feature, this issue of extrapolation disappears since the model only uses the remaining input features, which follow the original data distribution.

Recently, \cite{PFI_bias} proposed a decomposition of the PFI estimator to explain this extrapolation bias. In particular, they decompose the PFI into the standard bias-variance decomposition arising from the estimation of the model, plus an additional bias term which, in our case under the null hypothesis, is exactly zero. Indeed, this term corresponds to a cross term between the bias of the model under the perturbed distribution and the quantity $m(X') - m(X)$, where $m$ is the Bayes predictor. We observe that since $X$ and $X'$ coincide in all coordinates except the $j$-th one, and since the Bayes predictor is functionally independent of this covariate, this term vanishes. We observe that in all their experiments, this term is indeed zero for the null features.

Moreover, in Appendix D.2 of \cite{lobo2025doublerobustnessconditionalfeature}, they investigate empirically the convergence of PFI to zero for standard machine learning models. In this section, we extend their analysis by including more complex models such as the SuperLearner \citep{van2007super} and state-of-the-art tabular learning methods such as TabICL \citep{qu2026tabiclv2}, as well as by considering more challenging data-generating settings. In particular, we revisit their quadratic Gaussian setting with two features (\ref{subsub:gauss_quad}), extend it to non-Gaussian inputs (\ref{subsub:nongauss_quad}), higher-dimensional sparse linear models (\ref{subsub:sparse_lin}), non-sparse linear models (\ref{subsub:nonsparse_lin}), nonlinear high-dimensional models (\ref{subsub:sparse_nonlin}), and the Friedman benchmark (\ref{subsub:friedman}). We also consider classification settings with linear structure (\ref{subsub:lin_class}) and nonlinear interactions (\ref{subsub:compl_class}). Also, we study multiple correlation strengths across each setting. We use the cross-validated PFI from \texttt{HiDimStat} (\url{https://hidimstat.github.io}).

\subsection{Models}

We consider a collection of standard machine learning models for both regression and classification.

For regression, the base learners include:
\textit{lr} (Linear Regression), \textit{lasso} (Lasso regression), \textit{dt} (Decision Tree), \textit{rf} (Random Forest), \textit{et} (Extra Trees), \textit{gb} (Gradient Boosting), \textit{hgb} (Histogram-based Gradient Boosting), \textit{ab} (AdaBoost), \textit{bag} (Bagging), \textit{mlp} (Multilayer Perceptron), \textit{svr} (Support Vector Regression), and \textit{knn} (k-Nearest Neighbors).

We additionally consider an ensemble model, \textit{SuperLearner}, defined as a stacking regressor combining all base learners with a linear regression meta-learner. When applicable, we also include \textit{TabICL}, a recent deep learning model for tabular data based on in-context learning.

For classification, we use the analogous set of models:
\textit{lr} (Logistic Regression), \textit{dt} (Decision Tree), \textit{rf} (Random Forest), \textit{et} (Extra Trees), \textit{gb} (Gradient Boosting), \textit{hgb} (Histogram-based Gradient Boosting), \textit{ab} (AdaBoost), \textit{bag} (Bagging), \textit{mlp} (Multilayer Perceptron), \textit{svm} (Support Vector Machine), and \textit{knn} (k-Nearest Neighbors).

As in the regression case, we include a \textit{SuperLearner}, implemented as a stacking classifier with logistic regression as the meta-learner, as well as \textit{TabICL}, a deep tabular classification model based on in-context learning.

\subsection{Simulation settings}

We consider a collection of synthetic regression and classification benchmarks designed to evaluate feature importance methods under different data-generating mechanisms. In all settings, $\rho\in \{0.3, 0.6, 0.9\}$ denotes the correlation among predictors whenever applicable.

\subsubsection{Gaussian quadratic}\label{subsub:gauss_quad}

The input is 2-dimensional Gaussian $X \sim \mathcal{N}(0,\Sigma)$, 
with covariance matrix is $\Sigma_{i, j}=\rho^{|i-j|}$.

The response is generated as $Y=X_1^2+\varepsilon$, where $\varepsilon\sim\mathcal{N}(0,1)$. Only the first variable is truly relevant.

\subsubsection{Non-Gaussian quadratic}\label{subsub:nongauss_quad}

A Gaussian copula is used to generate correlated non-Gaussian predictors. First,
$
Z\sim\mathcal{N}(0,\Sigma),
$
with the same covariance matrix as above. The variables are transformed according to
\[
U_j=\Phi(Z_j), \qquad
X_j=F^{-1}_{\chi^2_3}(U_j),
\]
where $\Phi$ is the standard normal cumulative distribution function and
$F^{-1}_{\chi^2_3}$ is the quantile function of the $\chi^2$ distribution with
three degrees of freedom. The response is the same as before,
$
Y=X_1^2+\varepsilon,$
with $\varepsilon\sim\mathcal{N}(0,1)$. Again, only $X_1$ is relevant.

\subsubsection{Sparse high-dimensional linear regression}\label{subsub:sparse_lin}

The input dimension is $p=100$. Predictors follow
$
X\sim\mathcal{N}(0,\Sigma),
$
where the covariance matrix has autoregressive structure
$
\Sigma_{ij}=\rho^{|i-j|}.
$
The response linear is
$
Y=X^\top\beta+\varepsilon,
$
with $\varepsilon\sim\mathcal{N}(0,1)$. The coefficient vector satisfies
\[
\beta_j=
\begin{cases}
1, & j\in\{1,11,21,\ldots,91\},\\
0, & \text{otherwise}.
\end{cases}
\]
Hence, ten features are truly important.

\subsubsection{Non-sparse linear regression}\label{subsub:nonsparse_lin}

The data are generated as in the sparse linear model, except that the active variables are
\[
\{1,4,7,\ldots,97\},
\]
that is, every third predictor carries signal. This produces a substantially denser coefficient vector.

\subsubsection{Sparse nonlinear interactions}\label{subsub:sparse_nonlin}

The predictor dimension is again $p=100$ with autoregressive covariance
$
\Sigma_{ij}=\rho^{|i-j|}.
$
The response is
\[
\begin{aligned}
Y={}&X_1X_{11}
+X_{21}^2
+\sin(X_{31})\\
&+X_{41}X_{51}
+X_{61}^2
+\varepsilon,
\end{aligned}
\]
where $\varepsilon\sim\mathcal{N}(0,1)$.
The truly relevant variables are
\[
\{1,11,21,31,41,51,61\}.
\]

\subsubsection{Friedman benchmark}\label{subsub:friedman}

We use the classical Friedman~1 regression benchmark with
$p=10$ predictors and additive Gaussian noise of variance one. The response is
\[
Y
=
10\sin(\pi X_1X_2)
+
20(X_3-0.5)^2
+
10X_4
+
5X_5
+\varepsilon,
\]
where
$
X_j\stackrel{i.i.d.}{\sim}\mathrm{Uniform}(0,1),
$ and $
\varepsilon\sim\mathcal{N}(0,1).
$
Only the first five variables are relevant.

\subsubsection{Linear classification}\label{subsub:lin_class}

The predictor dimension is $p=20$ with autoregressive covariance
$
\Sigma_{ij}=\rho^{|i-j|}.
$
The linear predictor is
$
\eta=X^\top\beta,
$
where
\[
\beta_j=
\begin{cases}
(-1)^k, & j\in\{1,4,7,10,13,16,19\},\\
0, & \text{otherwise},
\end{cases}
\]
and $k$ indexes the active variables, yielding alternating positive and negative coefficients. The probability is
\[
P(Y=1\mid X)
=
\frac{1}{1+\exp(-\eta)},
\]
and
\[
Y\mid X\sim\mathrm{Bernoulli}(P(Y=1\mid X)).
\]

\subsubsection{Complex nonlinear classification}\label{subsub:compl_class}

The predictor dimension is again $p=20$ with autoregressive covariance structure. The linear predictor is
\[
\eta
=
2X_0
-
1.5X_1
+
X_2X_3
+
\sin(X_4),
\]
and the response follows
\[
P(Y=1\mid X)
=
\frac{1}{1+\exp(-\eta)},
\]
and
\[
Y\mid X\sim
\mathrm{Bernoulli}(P(Y=1\mid X)).
\]
The first five predictors are the only relevant variables.

\subsection{Asymptotic relevance in practice}

\begin{figure*}[htbp]
    \includegraphics[width=0.9\textwidth]{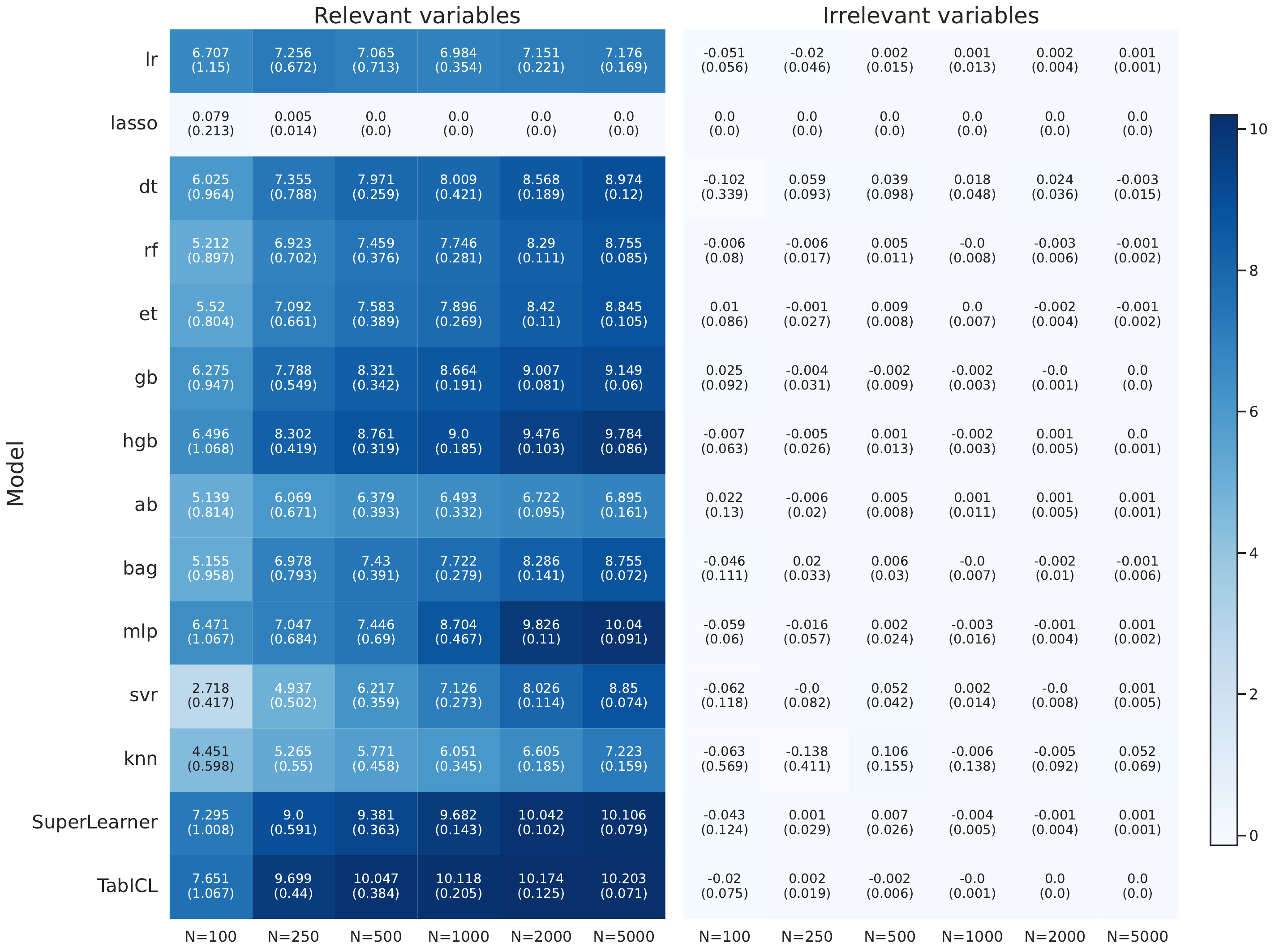}

    \caption{\textbf{Asymptotic relevance in the Friedman setting.} The left panel shows the mean cross-validated PFI across important features, while the right panel shows the corresponding values for unimportant features. As expected, null features receive essentially zero importance.}
    \label{fig:asympt_friedman}
\end{figure*}

\begin{figure*}[htbp]
        \includegraphics[width=0.9\textwidth]{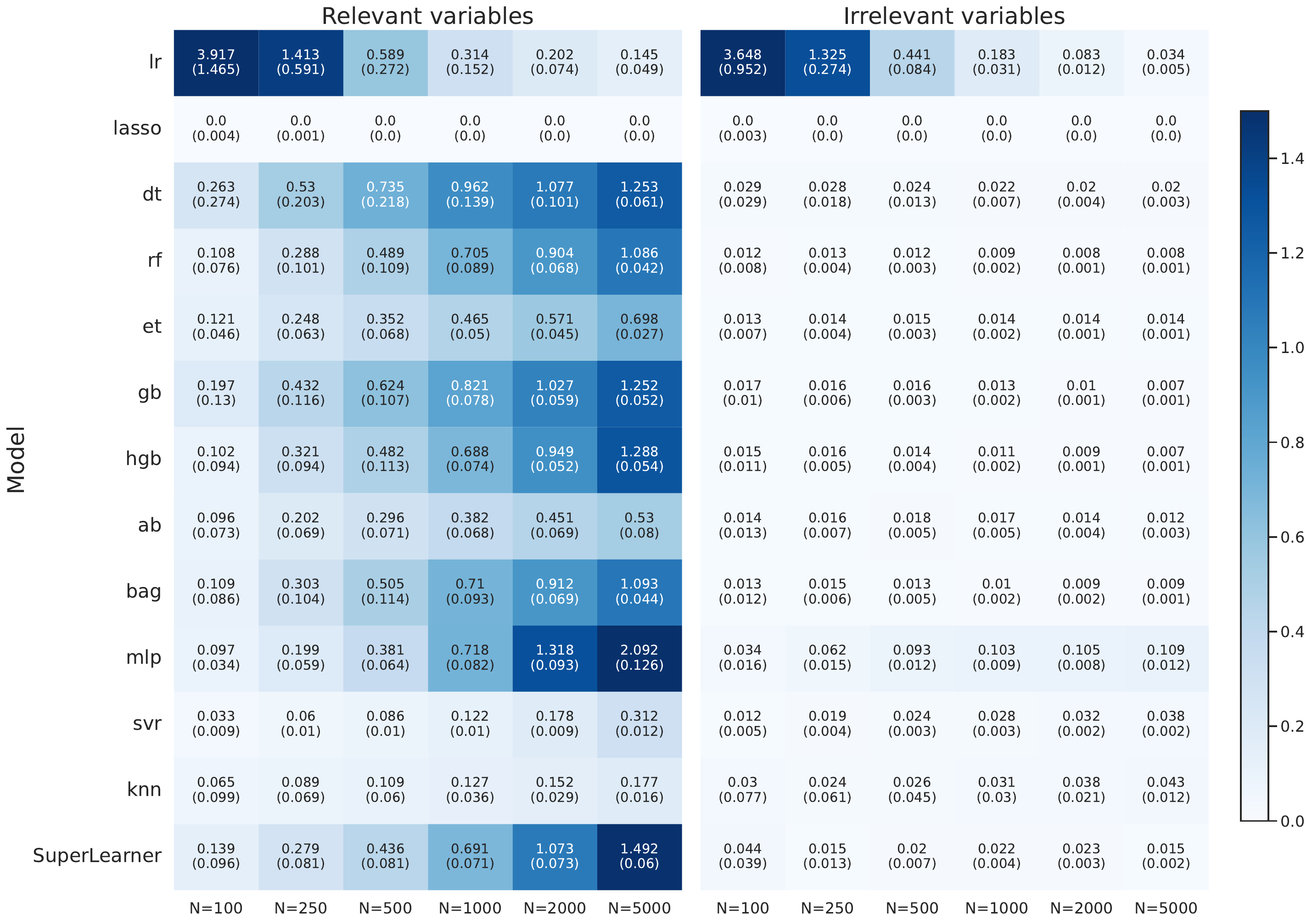}

    \caption{\textbf{Asymptotic relevance in high-dimensional sparse linear data with correlation 0.9:} The left panel shows the mean cross-validated PFI across important features, while the right panel shows the corresponding values for unimportant features. As expected, null features receive essentially zero importance.
}
    \label{fig:asympt_sparse_nonlin}
\end{figure*}

\begin{figure*}[htbp]
        \includegraphics[width=0.9\textwidth]{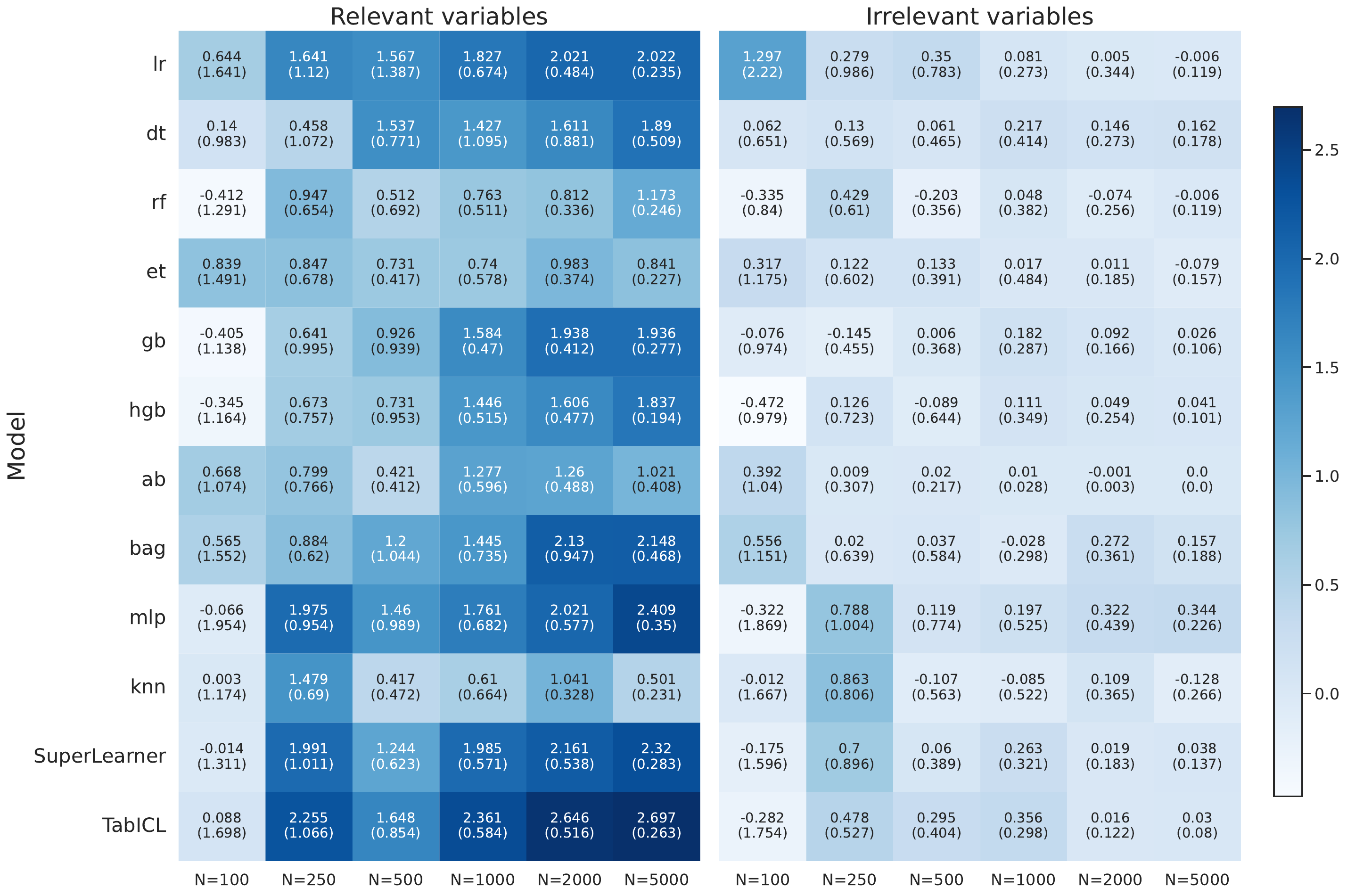}

    \caption{\textbf{Asymptotic relevance in classification with interactions setting with correlation 0.9:} The left panel shows the mean cross-validated PFI across important features, while the right panel shows the corresponding values for unimportant features. As expected, null features receive essentially zero importance.
}
    \label{fig:asympt_classif}
\end{figure*}

We evaluated all settings with feature correlations of $0.3$, $0.6$, and $0.9$, except for the Friedman setting, where the input distribution is fixed by construction. Since high correlation represents the most challenging setting, we report only a few illustrative examples for a correlation coefficient of $0.9$. We first present the Friedman setting in Figure~\ref{fig:asympt_friedman}, followed by a high-dimensional setting with sparse nonlinear interactions in Figure~\ref{fig:asympt_sparse_nonlin}, and a complex nonlinear classification setting in Figure~\ref{fig:asympt_classif}. For computational reasons, we do not report results for TabICL in the high-dimensional settings. Overall, we observe that all methods assign negligible importance to the null features, supporting the asymptotic relevance assumption. In some highly correlated settings, however, the MLP and KNN models assign a small positive importance to null features; nevertheless, these values remain substantially lower than those assigned to truly relevant features.


\section{Additional experiments}\label{sect:add_exp}

In this section, we present the remaining results on variable importance and additional convergence behaviors. This provides a more complete view of the experiments.

Note that for our simulated data, we did not replicate the exact setup from \cite{Ewald_2024}, in which the unimportant coordinates $X_1$ and $X_2$ are nearly identical. Indeed, this induces collinearity, which prevents accurate estimation of the linear model and results in a model that includes the term $0.36X_1 - 0.36X_2$.

As a consequence, \cite{Ewald_2024} argue that PFI incorrectly assigns importance to these uninformative covariates. However, this is actually an estimation issue, since the linear model is not consistent in cases of perfect correlation; in contrast, the theoretical PFI does not attribute any importance to such features. Furthermore, this setup violates the identifiability assumption, which requires that no coordinate be a deterministic function of the others.

\subsection{Simulated data}\label{app:sim_data}

\begin{figure*}[htbp]
        \includegraphics[width=0.9\textwidth]{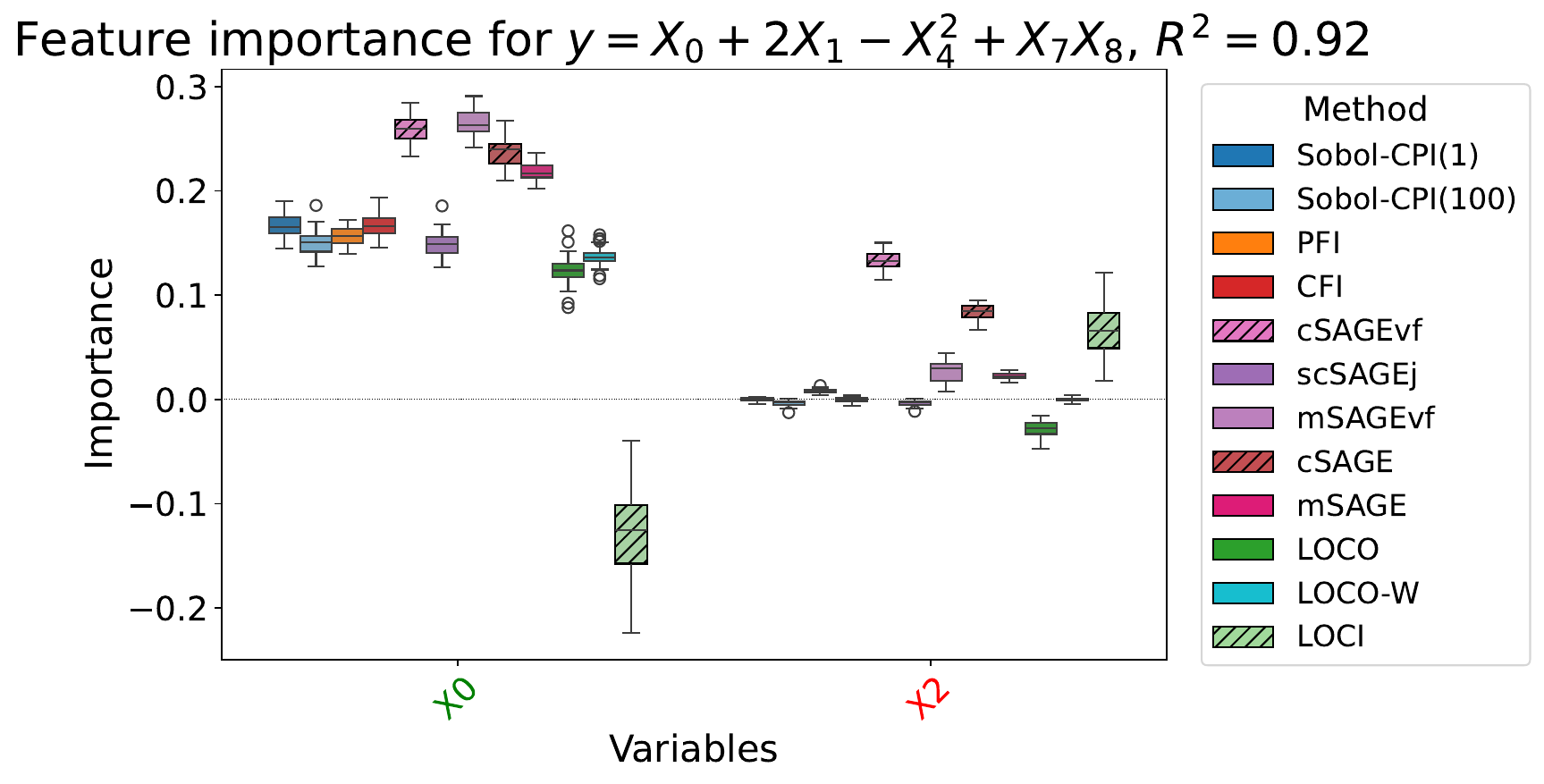}

    \caption{\textbf{Boxplot of estimated VIMs for an important and an unimportant variable using a Random Forest:} Methods not satisfying theoretically the minimal axiom have boxes filled with diagonal hatch lines. The left panel shows an important covariate, while the right panel shows an unimportant one. Only conditional SAGE(vf) and LOCI, which
fail to satisfy the minimal axiom, assign non-zero importance to the unimportant
variable. \texttt{Sobol-CPI(1)}, \texttt{Sobol-CPI(100)}, \texttt{CFI}, \texttt{scSAGEj}, \texttt{LOCO}, and \texttt{LOCO-W} aim to estimate $\psi_\mathrm{TSI}$.
}
    \label{fig:reduced_RF}
\end{figure*}

From Figure \ref{fig:reduced_RF}, where we used a Random Forest, we observe the same overall pattern as in Figure \ref{fig:reduced_boxplot}, where the main model is a Gradient Boosting machine. We also note that the Gradient Boosting model is more accurate (achieving a better $R^2$), but its estimated importance values exhibit higher variability. Therefore, even though both models are expected to converge to the same quantity, some inference methods may be more desirable than others.

For instance, Sobol-CPI (and similarly, CFI) benefits from a double robustness property \citep[see][]{lobo2025doublerobustnessconditionalfeature}, which enhances its ability to detect null covariates and leads to better performance on the null feature $X^2$ compared to PFI. Additionally, we observe that LOCI not only fails to satisfy the minimal axiom—by assigning importance to uninformative covariates—but also assigns negative importance  
to important covariates, exhibiting behavior that is clearly undesirable. This could indicate substantial model misspecification due to a large amount of unexplained variance. 

\begin{figure*}[htbp]
        \includegraphics[width=0.9\textwidth]{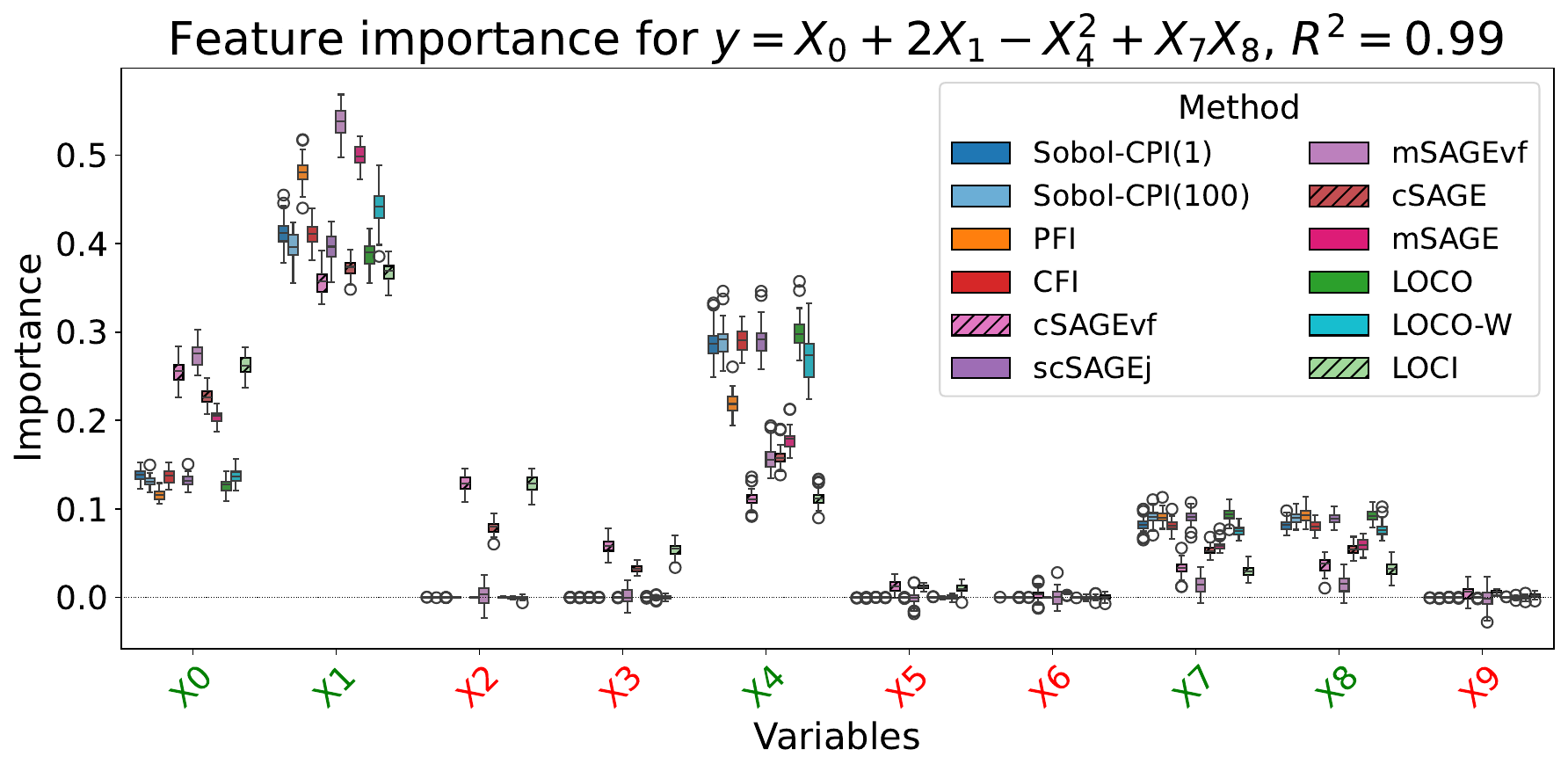}

    \caption{\textbf{Boxplot of estimated VIMs for all features using Gradient Boosting:} Methods not satisfying theoretically the minimal axiom have boxes filled with diagonal hatch lines. Important covariates are shown in green, and unimportant ones in red. Only conditional SAGE(vf) and LOCI, which
fail to satisfy the minimal axiom, assign non-zero importance to the unimportant
variable. \texttt{Sobol-CPI(1)}, \texttt{Sobol-CPI(100)}, \texttt{CFI}, \texttt{scSAGEj}, \texttt{LOCO}, and \texttt{LOCO-W} aim to estimate $\psi_\mathrm{TSI}$.
}
    \label{fig:complete_boxplot_GB}
\end{figure*}

\begin{figure*}[htbp]
        \includegraphics[width=0.9\textwidth]{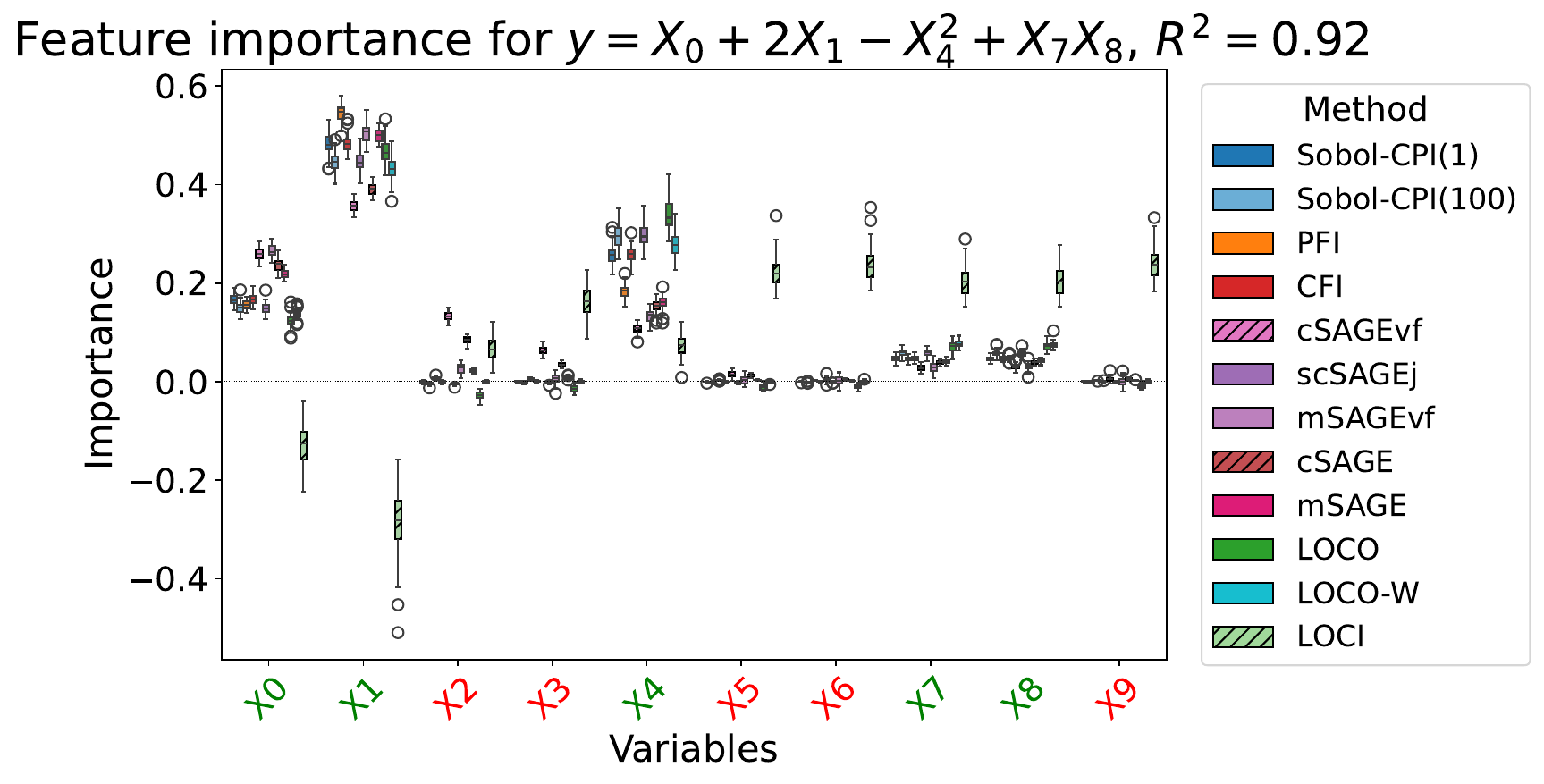}

    \caption{\textbf{Boxplot of estimated VIMs for all features using Random Forest:} Methods not satisfying theoretically the minimal axiom have boxes filled with diagonal hatch lines. Important covariates are shown in green, and unimportant ones in red. Only conditional SAGE(vf) and LOCI, which
fail to satisfy the minimal axiom, assign non-zero importance to the unimportant
variable. \texttt{Sobol-CPI(1)}, \texttt{Sobol-CPI(100)}, \texttt{CFI}, \texttt{scSAGEj}, \texttt{LOCO}, and \texttt{LOCO-W} aim to estimate $\psi_\mathrm{TSI}$.
}
    \label{fig:complete_boxplot_RF}
\end{figure*}

For completeness, in Figures \ref{fig:complete_boxplot_GB} and \ref{fig:complete_boxplot_RF} we present the estimated importance for all the features, using a Gradient Boosting and a Random Forest respectively. 
From these figures, we observe that the VIMs satisfying the minimal axiom do not assign any importance to the red (uninformative) features. We also note the equivalence between \texttt{Sobol-CPI(1)}, \texttt{Sobol-CPI(100)}, \texttt{CFI}, \texttt{scSAGEj}, \texttt{LOCO}, and \texttt{LOCO-W}, even though they rely on different estimation approaches and models.

From Figure~\ref{fig:complete_boxplot_RF}, we first observe that the estimation quality is slightly lower due to the reduced performance of the model. Notably, LOCI exhibits highly undesirable behavior: it not only assigns importance to uninformative covariates, violating the minimal axiom, but also assigns negative importance to important covariates. This is likely a consequence of the variability in the optimization process. Overall, Sobol-CPI demonstrates more reliable inference results \citep[see][]{paillard2025measuringvariableimportanceheterogeneous}.

\subsubsection{Convergence}\label{sect:exp_convergence}

\begin{figure*}[htbp]
        \includegraphics[width=0.9\textwidth]{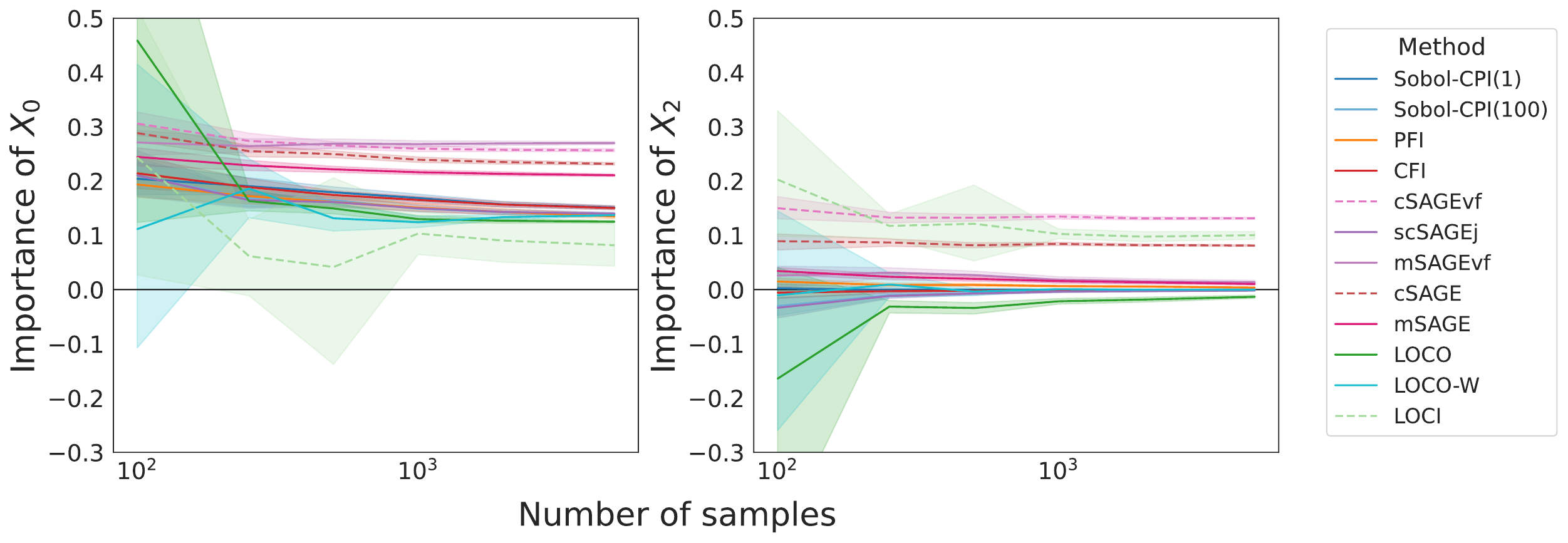}

    \caption{\textbf{Convergence of VIMs for an important and an unimportant covariate:} Methods not satisfying theoretically the minimal axiom have discontinuous lines. The left panel shows the convergence behavior of various VIMs for an important covariate, highlighting different trends as each method targets distinct indices with different objectives. The right panel illustrates the behavior for an unimportant covariate, where we can observe which VIMs satisfy the minimal axiom by assigning negligible or zero importance.
   }
    \label{fig:convergence}
\end{figure*}

In Figure \ref{fig:convergence}, we illustrate that the behavior observed in the boxplots from Figure \ref{fig:reduced_boxplot} is not specific to a fixed sample size $n$, but rather remains stable across different values of $n$. On the left panel, which corresponds to an important covariate, we observe that several distinct trends emerge early in the sampling process and remain consistently separated as $n$ increases. On the right panel, we observe that both SAGE and LOCI assign importance to features that are, in fact, not important. This suggests that these methods may fail to satisfy the minimal axiom, as they can attribute relevance to covariates that have no true impact on the outcome.

\begin{figure*}[htbp]
        \includegraphics[width=0.9\textwidth]{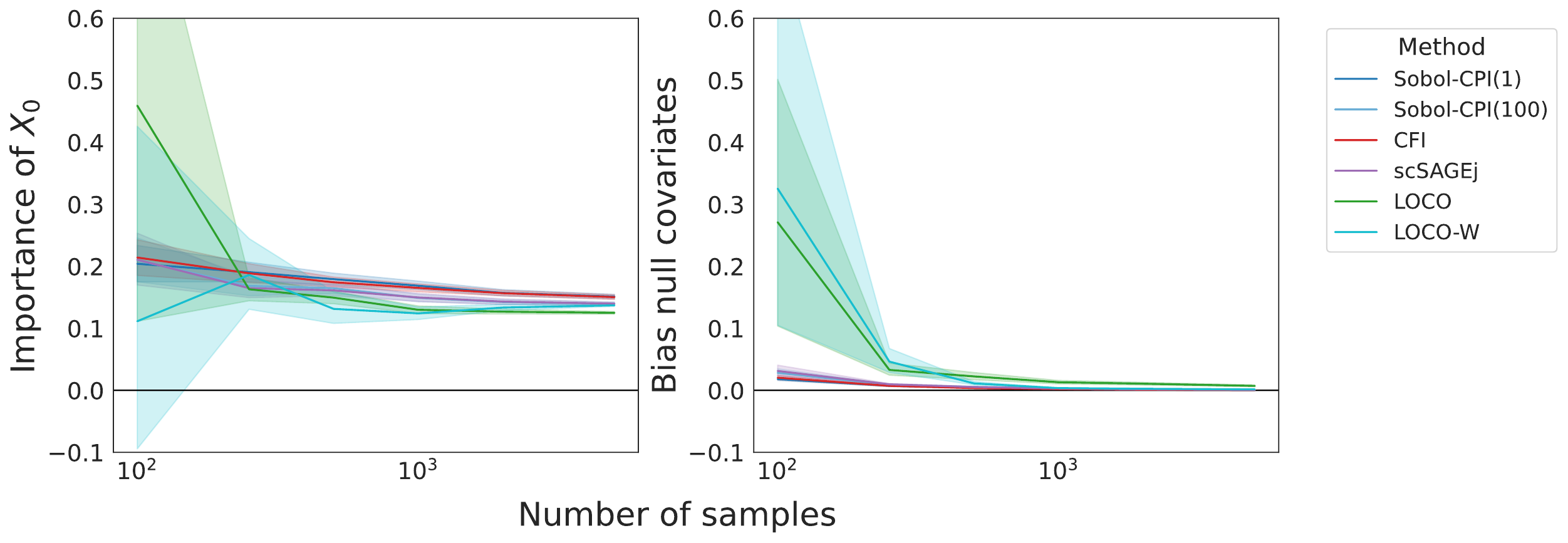}

    \caption{\textbf{Convergence of $\psi_\mathrm{TSI}$ estimates:} The left panel shows the convergence behavior of the $\psi_\mathrm{TSI}$ estimates for an important covariate. The right panel displays the mean bias across unimportant covariates.
}
    \label{fig:convergence_reduced}
\end{figure*}

In Figure~\ref{fig:convergence_reduced}, we focus exclusively on VIMs that target the total Sobol' index. As such, this figure does not aim to compare different theoretical importance indices, but rather to assess the inference properties of estimators targeting the same quantity—making it a purely methodological comparison. On the left, we observe that for the important covariate, the CPI-based methods exhibit lower variance compared to refitting-based approaches (\texttt{LOCO(-W)}), consistent with findings from \cite{paillard2025measuringvariableimportanceheterogeneous}. On the right, we report the mean bias across null covariates. Again, CPI-based methods show more favorable behavior due to their double robustness property, as discussed in \cite{lobo2025doublerobustnessconditionalfeature}.

\subsection{Real data}\label{app:bike}

\begin{figure*}[htbp]
        \includegraphics[width=0.99\textwidth]{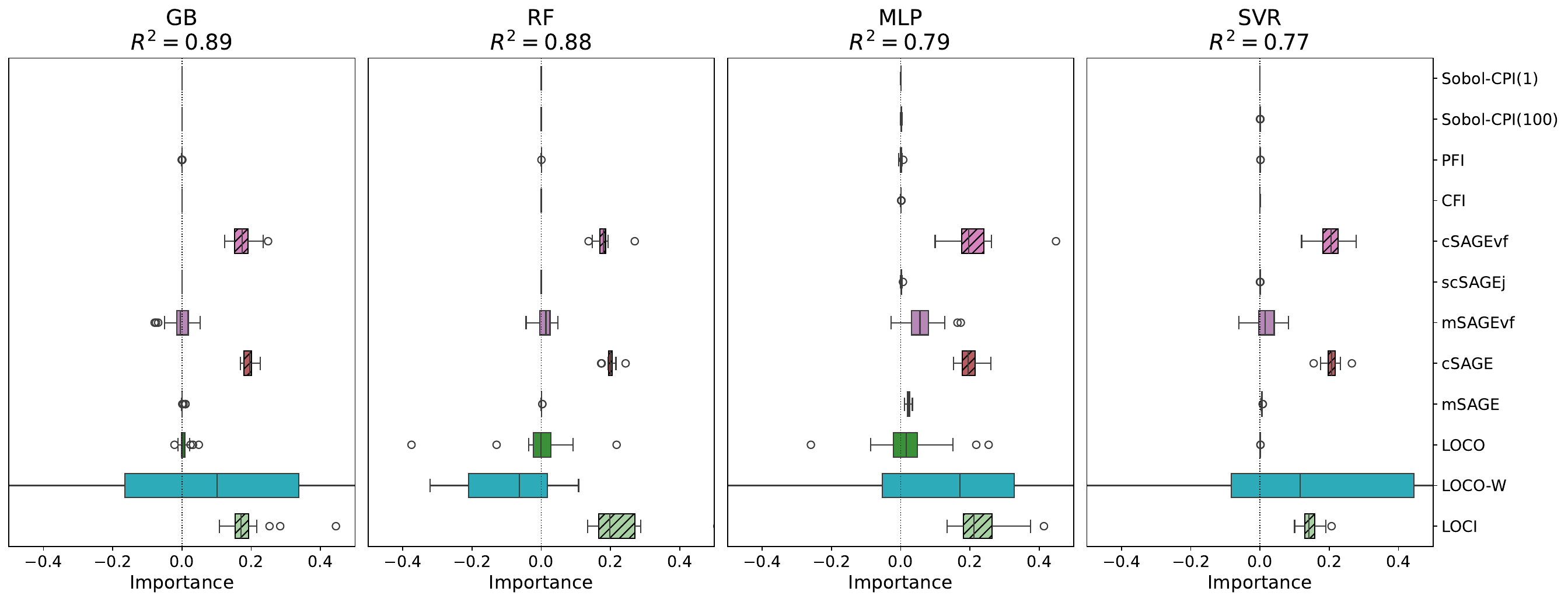}

    \caption{\textbf{Boxplots of the VIMs for the \textit{year} feature using Gradient Boosting, Random Forest, Neural Network and Support Vector Regression:} Methods not satisfying theoretically the minimal axiom have boxes filled with diagonal hatch lines. All methods that estimate an importance index satisfying the minimal axiom assign no importance to this variable.
}
    \label{fig:bike_year_all}
\end{figure*}

In this section, we replicate the results from the main text using the Bike dataset, replacing the Gradient Boosting model with alternative machine learning models. In Figure~\ref{fig:bike_year_all}, we reproduce the results from Figure~\ref{fig:bike_main_1feature} using a Random Forest, Neural Network, and SVR, respectively.

We plot the importance assigned to the feature \texttt{year} across the different methods. Overall, we observe that the methods that do not satisfy the minimal axiom, namely \texttt{cSAGEvf}, \texttt{cSAGE}, and \texttt{LOCI}, assign a non-zero importance to this feature, whereas the methods satisfying the minimal axiom correctly assign it an importance close to zero. In particular, we note that \texttt{PFI} successfully filters out the null feature. We also observe that the \texttt{LOCO} approach exhibits substantially higher variability due to the repeated model refitting. This variability is even more pronounced for \texttt{LOCO-W} because of the additional data-splitting step, which results in even less stable refitted models.

\begin{figure*}[htbp]
        \includegraphics[width=0.9\textwidth]{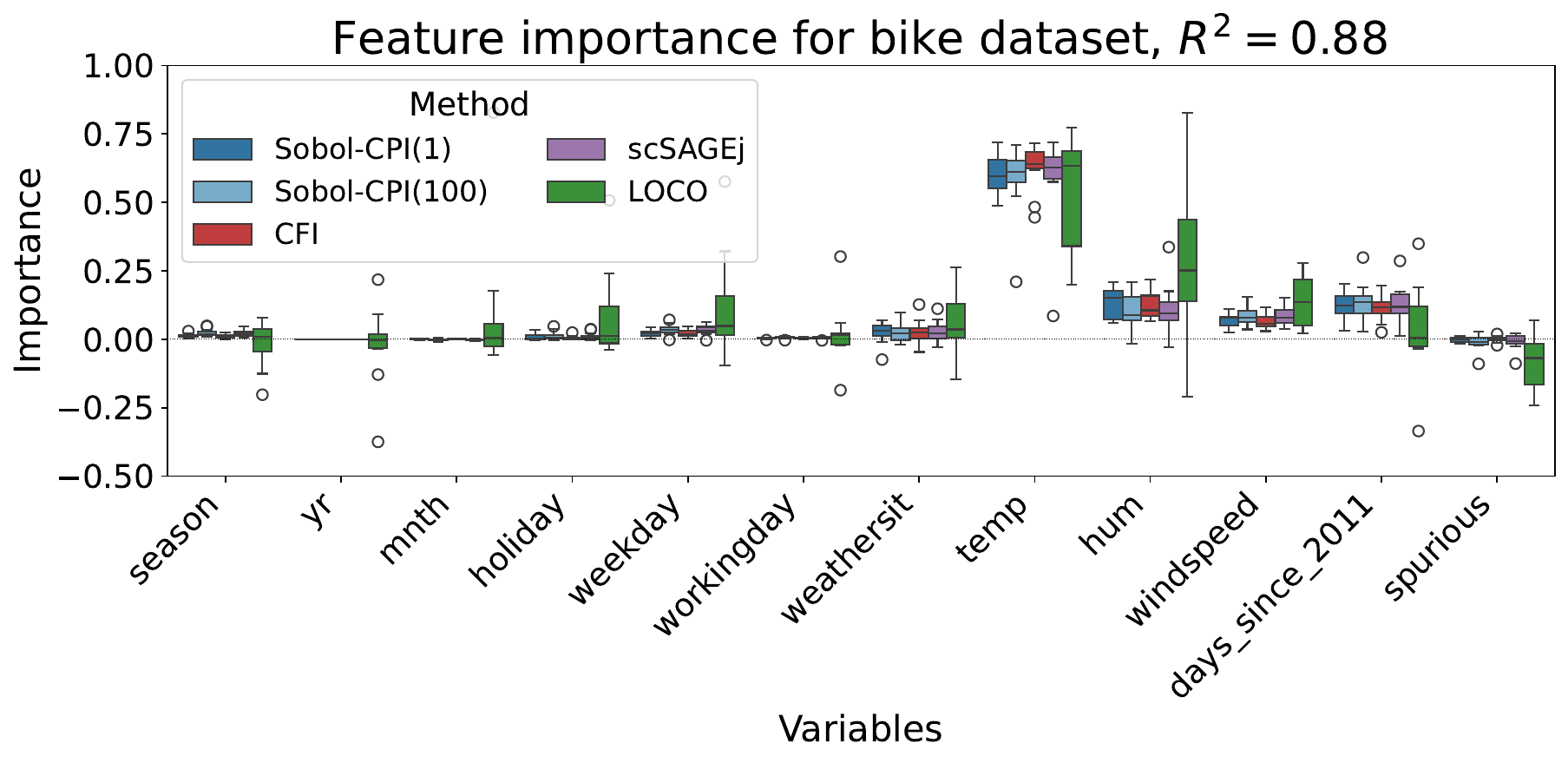}
    \caption{\textbf{Boxplots of the VIMs estimating $\psi_\mathrm{TSI}$ for all features using a Random Forest:} All methods aim to estimate the same theoretical quantity. While their estimates are generally close, the refitting-based approaches exhibit poorer inference properties.
}
    \label{fig:bike_TSI_RF}
\end{figure*}

\begin{figure*}[htbp]
        \includegraphics[width=0.9\textwidth]{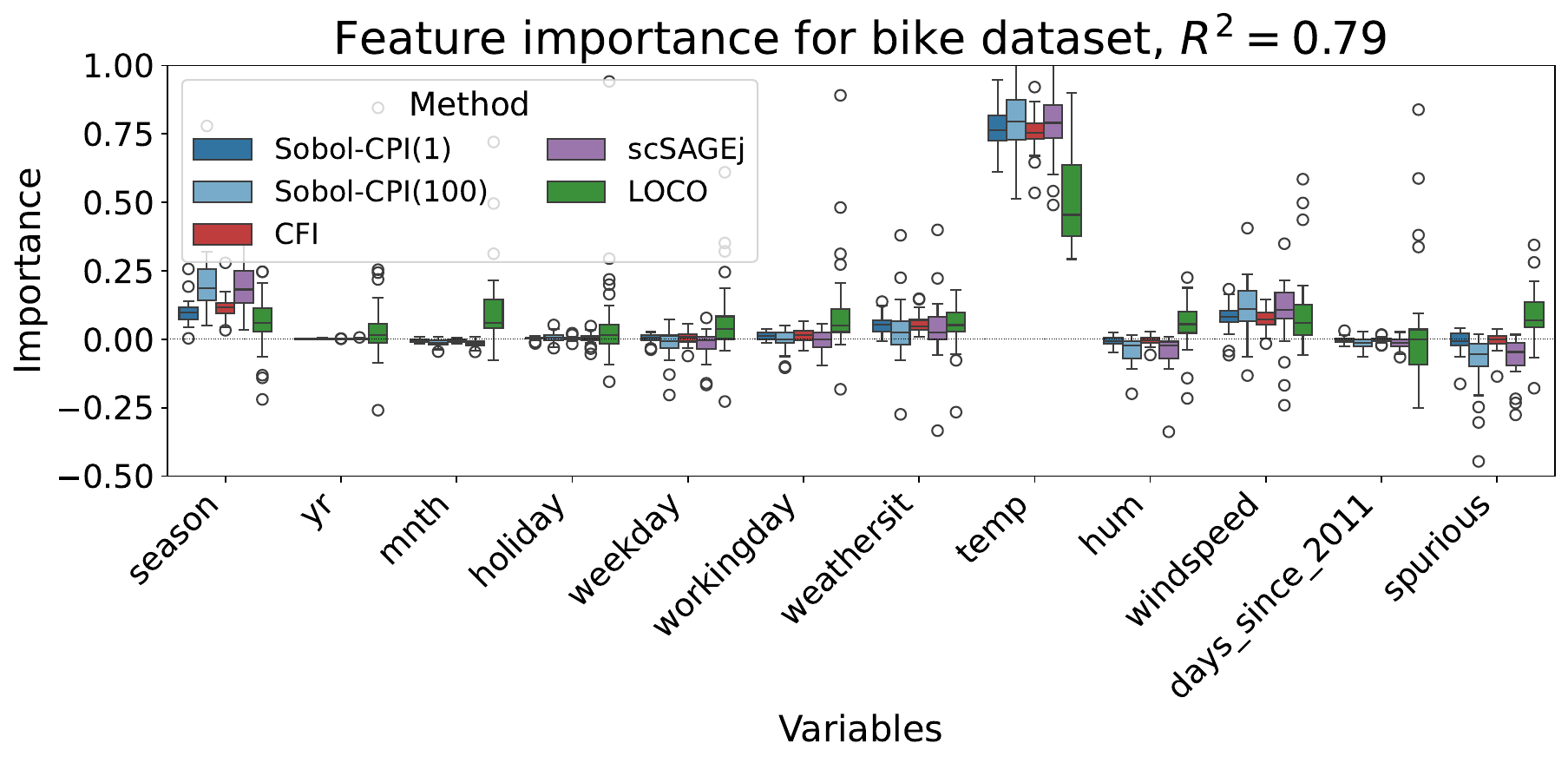}
    \caption{\textbf{Boxplots of the VIMs estimating $\psi_\mathrm{TSI}$ for all features using a Neural Network:} All methods aim to estimate the same theoretical quantity. While their estimates are generally close, the refitting-based approaches exhibit poorer inference properties.
}
    \label{fig:bike_TSI_NN}
\end{figure*}

\begin{figure*}[htbp]
        \includegraphics[width=0.9\textwidth]{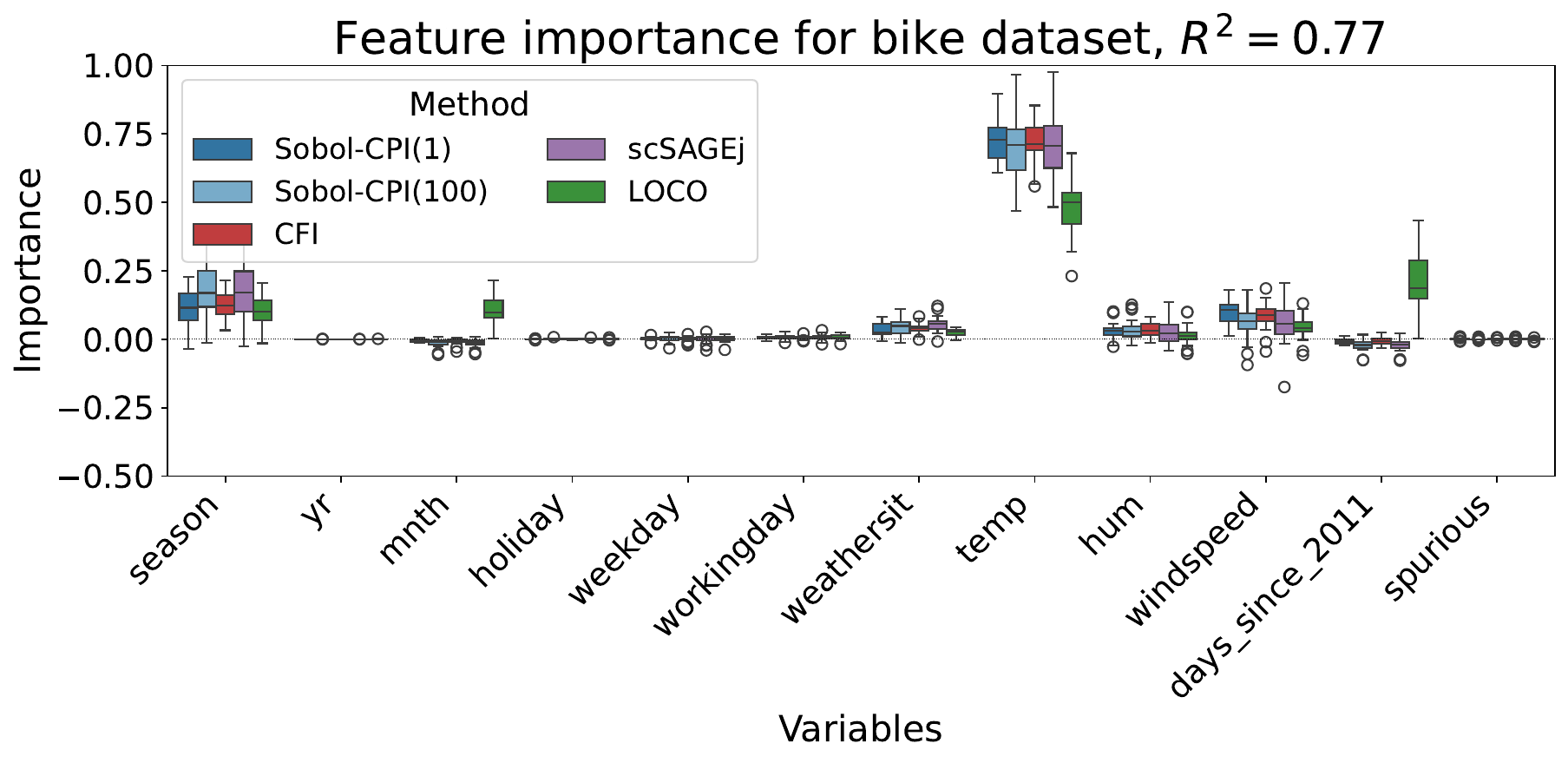}
    \caption{\textbf{Boxplots of the VIMs estimating $\psi_\mathrm{TSI}$ for all features using a SVR:} All methods aim to estimate the same theoretical quantity. While their estimates are generally close, the refitting-based approaches exhibit poorer inference properties.
}
    \label{fig:bike_TSI_SVR}
\end{figure*}

In Figures~\ref{fig:bike_TSI_RF}, \ref{fig:bike_TSI_NN}, and \ref{fig:bike_TSI_SVR}, we replicate the TSI estimation results shown in Figure~\ref{fig:bike_main_TSI} using respectively a Random Forest, Neural Network and SVR instead of Gradient Boosting. Overall, the conclusions remain unchanged: all methods target the same quantity, and consequently, their overall ranking is largely consistent. The main difference is that the LOCO approach, which relies on refitting the model, exhibits greater variability than the other methods, which are based on perturbation or marginalization. As a result, the LOCO estimates appear less stable.

\begin{figure*}[htbp]
        \includegraphics[width=0.99\textwidth]{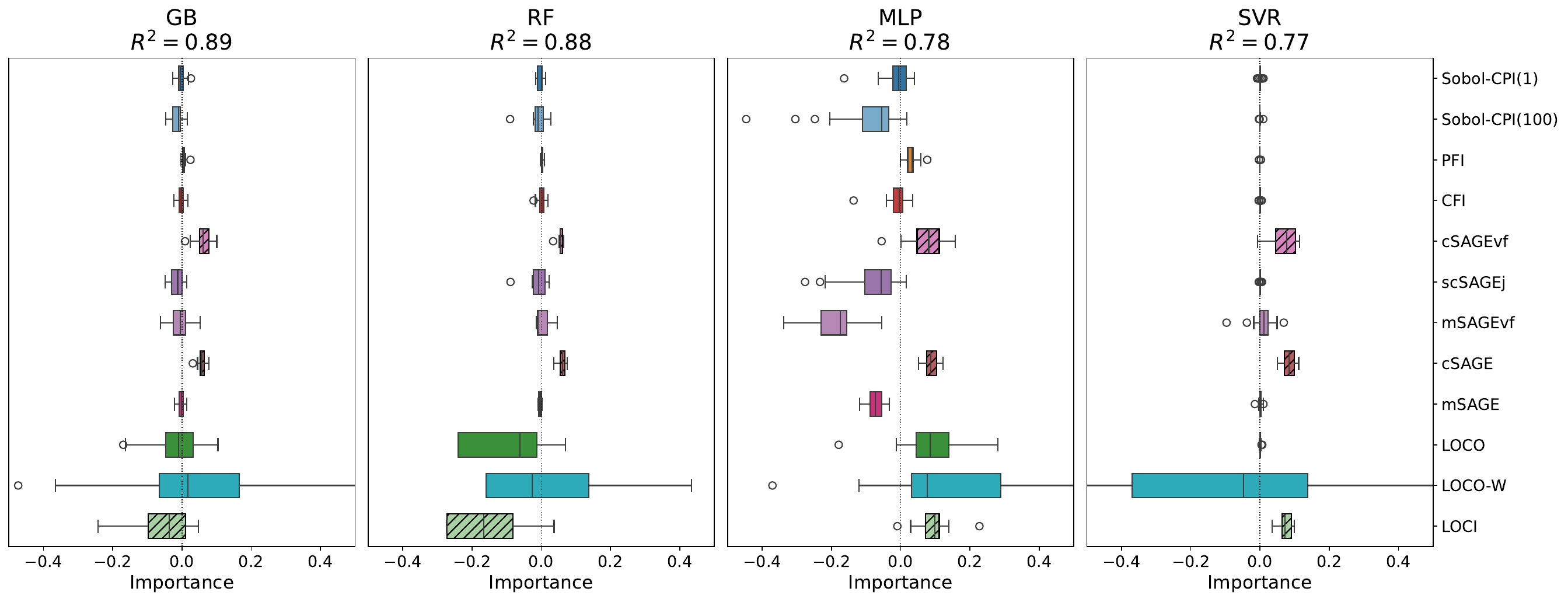}
    \caption{\textbf{Boxplots of the spurious feature importance in the Bike dataset:} Estimated importance across models and VIMs for a spurious feature with a correlation of 0.9 to a linear combination of the original features.
    Methods that do not satisfy the minimal axiom (hatched boxplots) generally assign non-zero importance to the spurious feature. 
}
    \label{fig:spurious_bike}
\end{figure*}

Nevertheless, the conclusion that the feature \texttt{year} is unimportant may not necessarily be correct, even though this conclusion is consistent across models and methods satisfying the minimal axiom. To distinguish between a genuinely unimportant feature and a truly null feature, we next consider an artificially constructed null feature. 
In Figure~\ref{fig:spurious_bike}, we present the estimated importance of an artificial null feature that has a correlation of 0.9 with a linear combination of the original features. We first observe that methods that do not satisfy the minimal axiom (hatched boxplots) generally assign non-zero importance to this feature, even though it is conditionally independent of the response. LOCI is a partial exception: while it occasionally exhibits high variability, it targets the same estimand as cSAGEvf and therefore also assigns a non-negligible importance.

Although LOCO asymptotically assigns zero importance to this feature, finite-sample estimation and optimization effects lead to highly variable estimates, which are often non-zero. Consequently, if the objective is variable selection, other TSI estimators may be preferable because they exhibit lower finite-sample variability.
We also note that PFI assigns zero importance to this null feature because the fitted model effectively filters it out during training.

\end{appendix}
\bibliographystyle{imsart-nameyear}


\bibliography{biblio} 

\end{document}